\documentclass{cas-dc}

\usepackage[authoryear,longnamesfirst]{natbib}
\usepackage{graphicx}
\usepackage{multirow}
\usepackage{todonotes}
\usepackage{float}
\usepackage{mathtools}
\usepackage{hyperref}
\usepackage{placeins}
\usepackage{lineno}

\def\tsc#1{\csdef{#1}{\textsc{\lowercase{#1}}\xspace}}
\tsc{WGM}
\tsc{QE}

\begin{document}
\let\WriteBookmarks\relax
\def\floatpagepagefraction{1}
\def\textpagefraction{.001}

\shorttitle{Singular Templates of Imaging Cherenkov Shower distribution (STOICS)}

\shortauthors{Shang, et al.}  

\title [mode = title]
{Singular Templates of Imaging Cherenkov Shower distribution (STOICS): A background estimation method for Very-High-Energy $\gamma$-ray observations}  



%

\author[1]{Ruo-Yu Shang}[orcid=0000-0002-9856-989X]

\cormark[1]

\fnmark[1]

\ead{r.y.shang@gmail.com}



\affiliation[1]{
            organization={Department of Physics and Astronomy, Barnard College, Columbia University},
            addressline={3009 Broadway}, 
            city={New York},
            postcode={10027}, 
            state={NY},
            country={USA}}

\author[2]{Karl Kosack}[orcid=0000-0001-8424-3621]

\fnmark[2]

\ead{karl.kosack@cea.fr}



\affiliation[2]{organization={Université Paris-Saclay, Université Paris Cité, CEA, CNRS, AIM},
            addressline={}, 
            city={Gif-sur-Yvette},
            postcode={91191}, 
            country={France}}

\author[3,4]{Brian Humensky}[orcid=0000-0002-1432-7771]

\fnmark[3]

\ead{humensky@umd.edu}



\affiliation[3]{organization={Department of Physics, University of Maryland},
            addressline={}, 
            city={College Park},
            state={MD},
            country={USA}}
\affiliation[4]{organization={NASA Goddard Space Flight Center},
            addressline={}, 
            city={Greenbelt},
            state={MD},
            country={USA}}

\cortext[1]{Ruo-Yu Shang}

\fntext[1]{}


\begin{abstract}
Analyses of Imaging Atmospheric Cherenkov Telescope (IACT) data for extended $\gamma$-ray sources face the issue that the field of view does not offer sufficient regions for background estimations.
In cases where the source angular size exceeds or occupies a significant part of the field of view, an independent background estimation method is necessary to carry out IACT analyses and to have a better understanding of the systematic uncertainties.
The proposed new method utilizes Singular Value Decomposition to extract the low-dimension representations of the distribution of cosmic-ray events in OFF runs and uses cosmic-ray-like events in the ON runs to estimate the background of $\gamma$-like events.
Using VERITAS archival data, we demonstrate that the new method is capable of providing reliable background modeling for observations across a wide range of observing conditions. 
\end{abstract}



\begin{keywords}
VHE $\gamma$ rays \sep 
Cosmic-ray \sep 
Background \sep
TeV-halos \sep
Extended sources \sep
\end{keywords}

\maketitle

\section{Introduction}
\label{sec:intro}

Extended $\gamma$-ray sources are ideal laboratories for studying particle acceleration mechanisms and transport. 
The extended sizes of these sources allow for detailed morphological analyses of these energetic objects.
Evolved pulsar wind nebulae (PWNe), supernova remnants (SNRs), and TeV halos are examples of such objects.
As an SNR-PWN system ages, the SNR reverse shock starts to interact with the PWN, creating an irregular nebula morphology, and the $\gamma$-ray emission is a sensitive probe to reveal the interactions that could accelerate the particles to the Tev-PeV regime.
A middle-aged ($t\sim 50$ kyr) SNR could have reached a size of $d\sim 80$ pc \citep{gelfand2009dynamical}.
Imaging Atmospheric Cherenkov Telescope (IACT) observatories with excellent $\gamma$-ray angular resolution (better than $0.08^{\circ}$ at $>1$ TeV) are ideal instruments for resolving the morphology details within these objects.

However, such a large angular size of an SNR exceeds or fills a significant part of the field-of-view of the current IACT observatories ($3.5^{\circ}-5.0^{\circ}$ for the current generation instruments), making it difficult to estimate the background of the $\gamma$-ray events.
These background events are events of cosmic-ray-induced air-showers mis-classified as $\gamma$-ray events.
Although it is possible to remove the majority of cosmic-ray events recorded by the IACT camera, an irreducible background of $\gamma$-like background events remains.
Conventionally, background estimation is achieved using outside sources in the field of view to normalize the background \citep{Berge_2007}, which is not available for extended source analysis.
New techniques utilizing external information for background estimation need to be developed in the cases of significantly extended sources.

Motivated by the need for background estimation methods for extended $\gamma$-ray sources, we present a novel technique, Singular Templates of Imaging Cherenkov Shower distribution (STOICS), which learns the correlations between the cosmic-ray-like events (i.e. events that failed the $\gamma$/hadron separation cut) and $\gamma$-ray-like background events in the OFF observations, and utilizes the cosmic-ray-like events from the ON observations to derive a background model for $\gamma$-ray-like background events of the ON observations.
In this paper, we review the $\gamma$-hadron separation variables in Section \ref{sec:gamma_hadron_separation} and the conventional methods in Section \ref{sec:review_methods}. 
Then we introduce the new method in Section \ref{sec:eigen_method}, and demonstrate the application of the method using the published VERITAS data in Section \ref{sec:example}.

\section{$\gamma$-hadron separation variables}
\label{sec:gamma_hadron_separation}

Imaging Atmospheric Cherenkov Telescopes (IACTs) detect particle cascades through Cherenkov radiation produced by charged particles traveling through the air.
The $\gamma$-ray-initiated showers and the cosmic-ray-initiated showers present intrinsically different shower shapes in the atmosphere due to the different particle interactions with the atmosphere.
The particles in $\gamma$-ray showers are produced in electromagnetic interactions, and their momenta are mostly aligned with the parent particles, resulting in narrow elliptic images of these showers when projected onto the camera of an IACT.
The particles in cosmic-ray showers are produced in hadronic interactions, the momenta of which can have significant transverse components due to decays of massive pions. Therefore, the images of cosmic-ray showers appear to be more irregular compared to those of $\gamma$-ray showers.

The Hillas parameters, which characterize the shower images as ellipses, are used for shower image parameterization \citep{Hillas_1985, Fegan_1997}.
The primary particle energy $E$, the shower impact parameter $R$ (the distance from the telescope to the shower axis), and the level of the night sky background (NSB) are important factors for the observed width and length of the ellipse. 
In order to minimize the dependency on these factors, the length and width of a shower image are compared against a set of $\gamma$-ray shower simulations at a given interval of shower size $s$ (the summation of the charge of all pixels of the image) and the impact parameter $R$ \citep{HESS2006Crab,Veritas_ghadron_separation_2006}. 
The derived parameters are mean reduced scaled length ($MSCL$)
\begin{equation}
    MSCL = \frac{1}{N} \sum_{i=1}^{N}
    \frac{l_{\mathrm{obs},i}(s,R)-\bar{l}_{\mathrm{sim}}(s,R)}{\sigma_{l,\mathrm{sim}}(s,R)},
\end{equation}
and mean reduced scaled width ($MSCW$)
\begin{equation}
    MSCW = \frac{1}{N} \sum_{i=1}^{N}
    \frac{w_{\mathrm{obs},i}(s,R)-\bar{w}_{\mathrm{sim}}(s,R)}{\sigma_{w,\mathrm{sim}}(s,R)},
\end{equation}
where N is the number of bright images detected by the IACT array, $l_{\mathrm{obs},i}$ ($w_{\mathrm{obs},i}$) is the shower length (width) observed by the $i$th telescope, $\bar{l}_{sim}$ ($\bar{w}_{sim}$) is the expected length (width) of the simulated events, and $\sigma_{l,\mathrm{sim}}$ ($\sigma_{w,\mathrm{sim}}$) is the 90\% containment variation of the length and width of the simulated events.
In this paper, we refer to events that pass the mean reduced scaled parameter cuts ($\gamma$-ray-like region) as $\gamma$-like events, and events that fail the mean reduced scaled parameter cuts (cosmic-ray-like region) as CR-like events.

Furthermore, modern IACT analyses often use multivariate machine-learning methods for $\gamma$/hadron separation. 
For example, the Random Forests (RF) and Boost-Decision Tree (BDT) methods take input of the Hillas parameters and stereoscopic parameters and improve $\gamma$/hadron separation performance by their better handling of parameter correlations \citep{krause2017improved}.
In this paper, we will not consider advanced machine-learning classifiers, and we will use the classical Hillas parameters for $\gamma$/hadron separation for simplicity.

In addition to the Hillas (or the multivariate) parameters of a shower image, the arrival direction of the shower, which is reconstructed by tracing the projected shower axis to the origin of the $\gamma$ ray, is an effective parameter to discriminate against isotropic cosmic-ray events as the $\gamma$ rays are charge-neutral and cannot be deflected by magnetic fields.

\section{Review of conventional background methods}
\label{sec:review_methods}

The background to the $\gamma$-ray events is cosmic-ray (CR)-induced air showers, which can be largely rejected in the analyses using the shower-image parameters described in Section \ref{sec:gamma_hadron_separation}.
The background events for the IACT analyses are the remaining cosmic-ray events that pass $\gamma$/hadron separation cuts and fall into the $\gamma$-like region.
 Background events mimic gamma-ray events because they originate from $\pi^{0}$ decays, which initiate electromagnetic cascades indistinguishable from those of primary $\gamma$ rays. 
The irreducible background cannot be removed despite further improvements in $\gamma$/hadron separation \citep{maier2007cosmic,sobczynska2009background}.

Given a number of counts $N_{\mathrm{ON}}$ in the ON region, the expected signal count $S$ is defined as
\begin{equation}
    S(\vec{r}) = N_{\mathrm{ON}}(\vec{r}) - \alpha N_{\mathrm{OFF}}(\vec{r}),
\end{equation}
where $\vec{r}$ is the sky coordinate of the region, $N_{\mathrm{OFF}}(\vec{r})$ is the background acceptance model sampled from a background control region (OFF region) and $\alpha$ is the background normalization factor.
The construction of the background for the IACT data can be divided into two aspects: (1) determination of the background acceptance model $N_{\mathrm{OFF}}(\vec{r})$ and (2) adjustment of the model to the data in the ON region.
Most conventional background methods, including the Ring Background, Reflected Region Background \citep{WobbleMode2001,HESS2006Crab,vovk2018spatial}, Field-of-View Background, and Template Background \citep{Rowell_Template2003,fernandes2014new}, use a background control region that is separated from the source region by an angular distance in the field of view to normalize the background acceptance function.
See also \citep{Berge_2007} for a review of the methods.

As an example, the Template Background constructs the background acceptance function $N_{\mathrm{OFF}}(\vec{r})$ using the events from the CR-like region.
The normalization $\alpha$ for the acceptance function is calculated as the number of $\gamma$-like events divided by the number of CR-like events in the camera field of view excluding the source region.
An advantage of the Template Background is that the background acceptance function is derived from the ON regions, therefore the localized observing effects, such as bright stars, will be effectively reduced.
The caveats of the Template Background, and most other conventional methods are that a sufficient event count is required in the field of view excluding the source region to calculate the $\alpha$ factor. 
If the source extension is large, the available background control region could have limited background statistics and lead to large uncertainties in the normalization factor $\alpha$ as well as in the background acceptance function.

In the cases of extended sources with angular extensions approaching the size of the field-of-view of the IACT observatories ($3.5^{\circ}-5.0^{\circ}$ diameter), because of the lack of off-source regions, there are insufficient events in the field of view for the conventional methods to derive the normalization.
Thus, the normalization factor $\alpha$ needs to be calculated using information outside the field of view.
For example, Holler \textit{et al.} \citep{holler2025utilizing} calculated the IACT detector response based on dedicated simulations of individual observations.

A classical approach to background estimation is the ON/OFF method, which takes an equal–length OFF run observation of a $\gamma$-ray-free field after the ON run at an equal zenith angle.
The ON/OFF method was applied for single telescope instruments, e.g. \citep{weekes1989observation}.
However, a drawback of this method is that it requires additional time to observe a $\gamma$-ray-free sky, thus losing observing time on the source.

An alternative is to use archival OFF runs from the past observations containing known point-like or small extension $\gamma$-ray sources under similar observing conditions (e.g. elevation, azimuth, night-sky brightness, etc.) that match the ON runs.
These OFF runs are assembled to construct a background template, and the template is normalized to the exposure of the ON runs.
For example, Mohrmann \textit{et al.} \citep{Mohrmann_2019} further refined the normalization calculation for OFF runs by taking into account the effects of observational elevation, optical transparency of the atmosphere, and throughput efficiency.
See also Wach \textit{et al.} \citep{wach2024background} for a similar approach.

Although the observing conditions are unique for every run, the $\gamma$-like events and the CR-like events collected from the same run are affected by the same observing condition.
Thus, systematic effects in $\gamma$-like events can be properly modeled by CR-like events if the information of the observing condition that is contained in the CR-like region can be extrapolated into the $\gamma$-like region.
The task is to find the parametrization for the correlation between CR-like events and $\gamma$-like background events that allows extrapolation.

\section{STOICS: A singular-template method for background modeling}
\label{sec:eigen_method}

\begin{figure}
\centering
\includegraphics[width=1.0\linewidth]{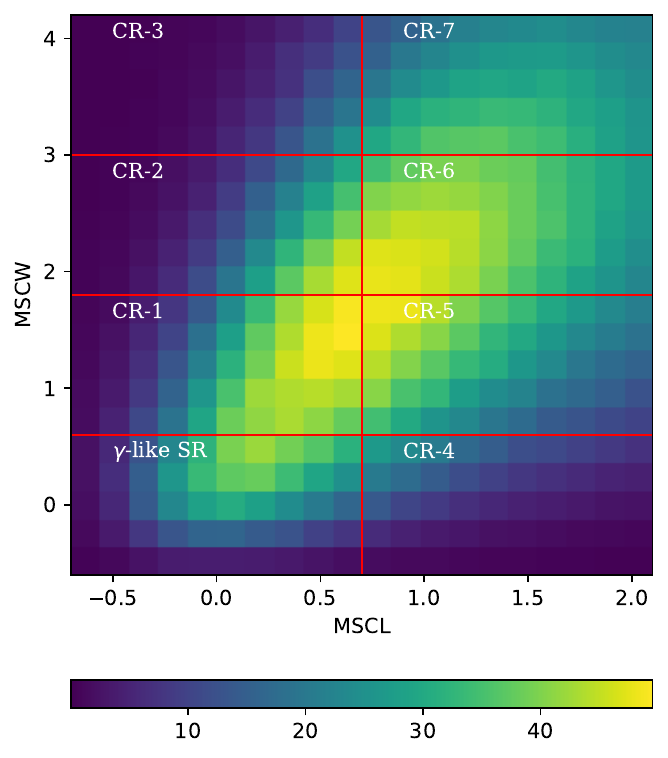}
\caption{The cosmic-ray event distributions in the space of $MSCL$ and $MSCW$ collected from VERITAS data of multiple blank-sky (no known $\gamma$-ray source) observations. The space is divided into eight analysis regions, with the $\gamma$-like signal region (SR) at the lower left corner containing events that pass the $\gamma$/hadron cut and seven cosmic-ray-like regions (CRs) containing events that fail the cut.
The color scale shows the count of events in each bin of the histogram.}
\label{fig:analysis_regions}
\end{figure}

The goal of the STOICS method is to predict the cosmic-ray background distribution in the $\gamma$-ray camera frame $(X,Y)$ in the $\gamma$-like signal region in the cases where the $\gamma$-ray source fills up the entire camera field of view.
The $\gamma$-like signal region is defined by the $\gamma$/hadron separation cut, $MSCL<MSCL_{\mathrm{cut}}$ and $MSCW<MSCW_{\mathrm{cut}}$, where $MSCL_{\mathrm{cut}}$ and $MSCW_{\mathrm{cut}}$ are the $\gamma$/hadron separation cut values.
In this study, a moderate cut of $MSCL_{\mathrm{cut}}=0.7$ and $MSCW_{\mathrm{cut}}=0.6$, which are typical cut values in the VERITAS analyses, is chosen.
The cosmic-ray events in the CR-like regions are used as control samples to make prediction for the $\gamma$-like region.
The definitions of these control regions are shown in Figure \ref{fig:analysis_regions}.
The control regions are chosen such that the majority of the cosmic-ray events is contained and the control regions have the same parameter-space size as the $\gamma$-like region.
We should note here that the STOICS method is not limited to analyses using $MSCL$ and $MSCW$ variables but can also be applied to analyses using a single multivariant variable (e.g., boost-decision tree) for $\gamma$/hadron separation.

\subsection{The profiles and the matrix of cosmic-ray templates}

\begin{figure*}
\centering
\includegraphics[width=1.0\linewidth]{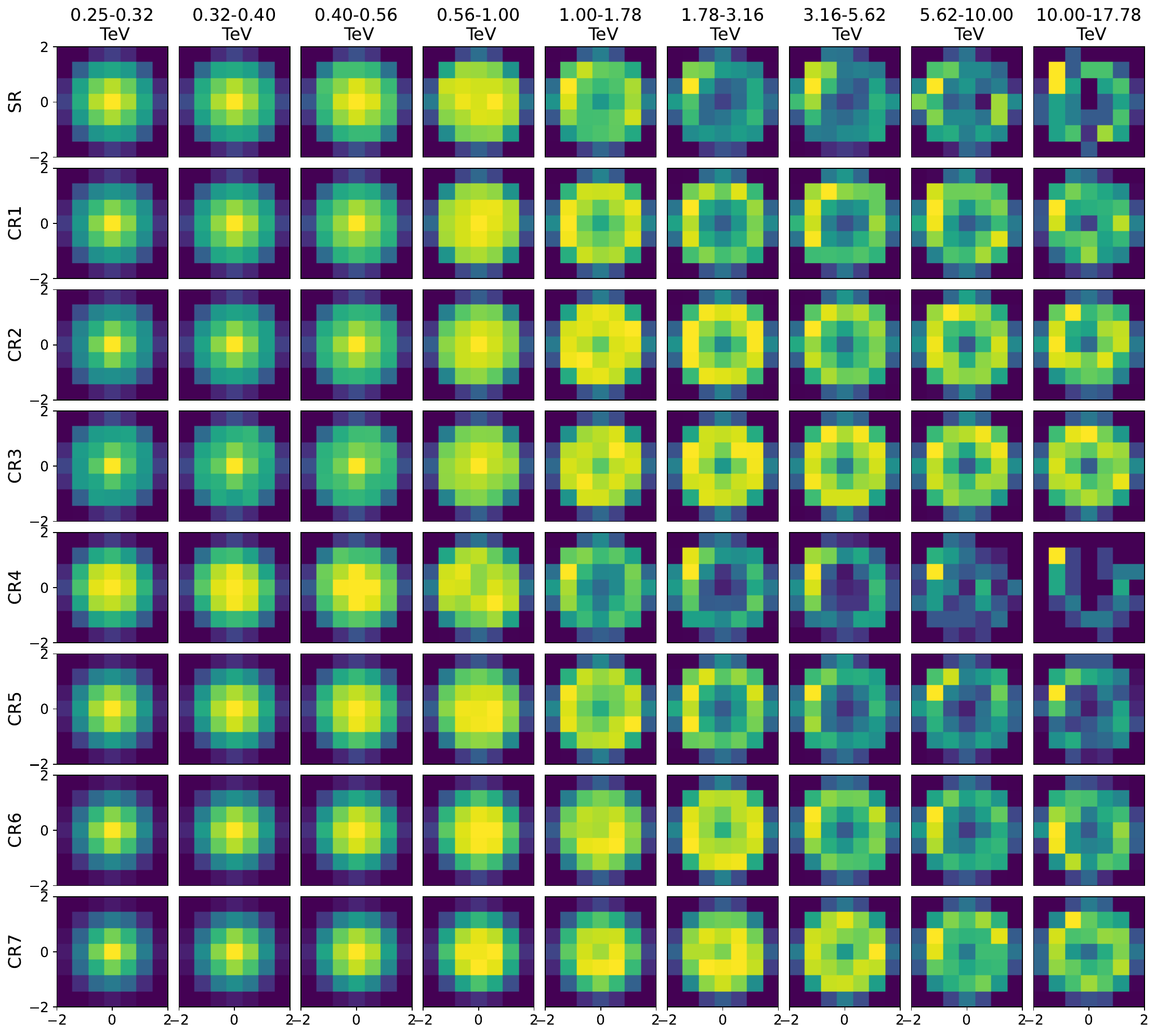}
\caption{Exemplary cosmic-ray event distributions in the camera frame ($2^{\circ}\times 2^{\circ}$) in the $\gamma$-like region and in the CR-like regions in different energy ranges.
There exist correlations between the background shapes in the SRs and the background shapes in the CRs.
The color scale shows the count of events in each bin of a mini-map. The color scales are not the same across different mini-maps and are saturated for the bin containing the largest count in a mini-map.}
\label{fig:background_camera_frame}
\end{figure*}

Figure \ref{fig:background_camera_frame} shows an array of energy-dependent 2-dimensional distributions of cosmic-ray counts in the $\gamma$-like region (SR) and the seven CR-like regions (CR$n$) in the camera $(X,Y)$ space for the energy range from 0.25 TeV to 17.78 TeV.
The distributions are binned into $N_{\mathrm{pix}} \times N_{\mathrm{pix}}$ pixels in the camera $(X,Y)$ space, and the choice of $N_{\mathrm{pix}}$ will be discussed in the following section. 
The pixels of the cosmic-ray event distributions are serialized into a single-column vector $\vec{x}$ for each observation run.
This single-column vector is referenced as a \textit{profile} in the following discussion.
All events used in this paper are selected from observations taken in good weather, with a minimum elevation angle of $40^{\circ}$ and passing basic reconstruction quality selection.

We then searched the VERITAS database to find $\gamma$-ray-free data whose observing conditions are matched to the observations of the source of interest.
The matched $\gamma$-ray-free data need to satisfy these criteria,
\begin{itemize}
\item Galactic $|b| > 10^{\circ}$ (only extragalactic fields)
\item
$\Delta_{\mathrm{El}} = |\theta_{\text{El}}^{\text{ON}}-\theta_{\text{El}}^{\text{OFF}}|<10^{\circ}$
\item $\Delta_{\mathrm{Az}} = |\cos \phi_{\text{Az}}^{\text{ON}}-\cos \phi_{\text{Az}}^{\text{OFF}}|<0.2$
\item $\Delta_{\mathrm{NSB}} = |\sigma_{\mathrm{ped}}^{\text{ON}}-\sigma_{\mathrm{ped}}^{\text{OFF}}|<2.0$ D.C.
\item Pointings of OFF observations have to be $>10^{\circ}$ away from the source of interest
\end{itemize}
, where $\theta_{\text{El}}$ is the averaged telescope pointing elevation during the observation, $\phi_{\text{Az}}$ is the averaged telescope pointing azimuth, and $\sigma_{\mathrm{ped}}$ is the night-sky brightness measured in the pedestal variance in the unit of digital count.
The source of interest observation parameters are labeled by a superscript `ON' while the matched $\gamma$-ray-free observation parameters are labeled by a superscript `OFF'.

The profiles $\vec{x}$ of the matched OFF runs are compiled into a template matrix $M$, each row of the matrix being the profile $\vec{x}_{i}$ of the $i$-th OFF run. For example, an element of the template matrix $M_{ij}$ represents the event count in the $j$-th entry of the profile vector in the $i$-th observation run.
If an element in $\vec{x}$ of an OFF run contains a known $\gamma$-ray source, this element is replaced by the `mirroring' element (located in the opposite location in the camera frame across the optical axis) from the same OFF run.
A typical matrix might have 3528 columns (8 regions $\times$ 9 energy bins $\times$ 49 spatial pixels) and many 10’s to 100’s of rows.

As we observe in Figure \ref{fig:background_camera_frame}, there exist correlations between the cosmic-ray event distributions (background acceptance) in the $\gamma$-like regions (SRs) and those in the cosmic-ray-like regions (CRs).
For example, low-energy background acceptance functions in SRs show a higher response at the center of the camera FoV while the high-energy ($>1$ TeV) background acceptance functions show a higher response around the edge of the camera FoV. A similar trend is also observed in the CRs.
Thus, it is desirable to develop an efficient algorithm that exploits the apparent correlations between the different event regions in the OFF observations to be used to predict the underlying background in the $\gamma$-like region of the ON observation from the events in the background regions of the latter.

Singular value decomposition (SVD) is a powerful tool to reduce data dimensionality and to extract correlations between different regions.
Applications of SVD in high-energy $\gamma$-ray astronomy can be reviewed in \cite{danaher1993application}.
We use SVD to factorize the template matrix, 
\begin{equation}
M
=
\begin{bmatrix}
\vec{x}_{0}^{\top} \\
\vec{x}_{1}^{\top} \\
\vec{x}_{2}^{\top} \\
\vdots
\end{bmatrix}
=
\sum_{k}^{r} \sigma_{k}
\vec{u}_{k} \vec{v}_{k}^{\top},
\end{equation}
where $r$ is the rank of $M$, $\vec{u}_{k}$ and $\vec{v}_{k}$ are the left and right singular vectors of $M$, and $\sigma_{k}$ are the singular values.
The singular vectors form a basis and each run profile $\vec{x}$ can be reconstructed as a linear combination of the singular vectors.
In this study, we required five OFF runs for each ON run.
The number of OFF runs per ON run is not strict, but a larger number ($\geq 2$) of OFF runs is preferred as it would ensure that the statistical noises in the singular vectors remain subdominant compared to the ones in the ON data profile vectors.

\begin{figure}
\centering
\includegraphics[width=1.0\linewidth]{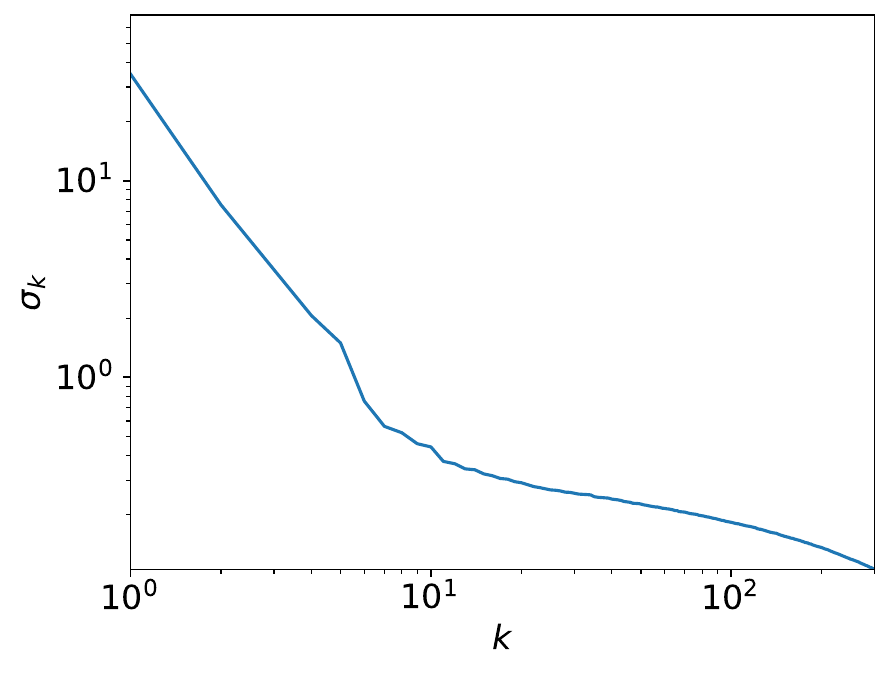}
\caption{Exemplary singular value $\sigma_{k}$ of a profile matrix compiled of a list of $\gamma$-ray-free runs in VERITAS database.}
\label{fig:singular_values}
\end{figure}

\begin{figure*}
\centering
\includegraphics[width=0.70\linewidth]{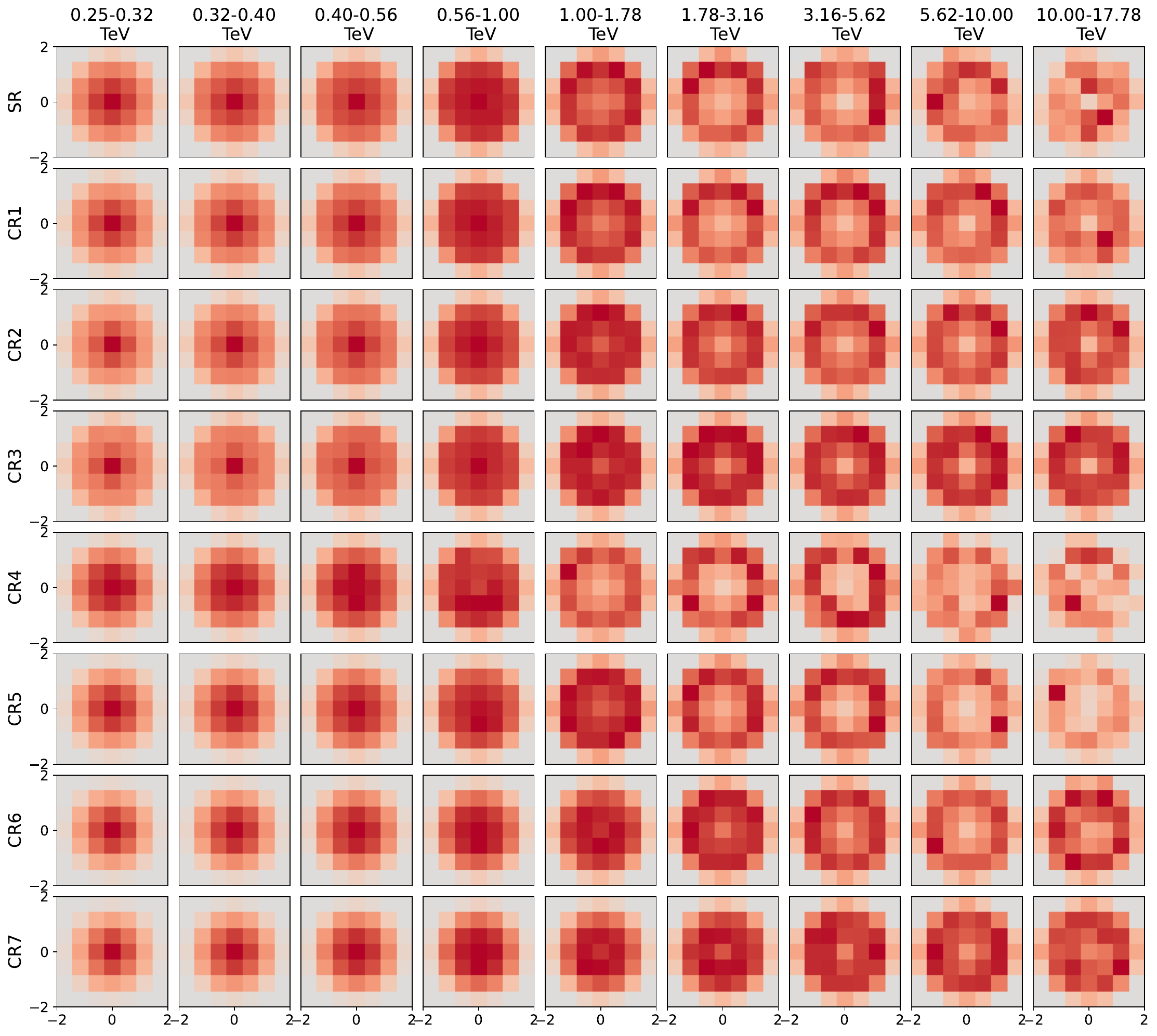}
\includegraphics[width=0.70\linewidth]{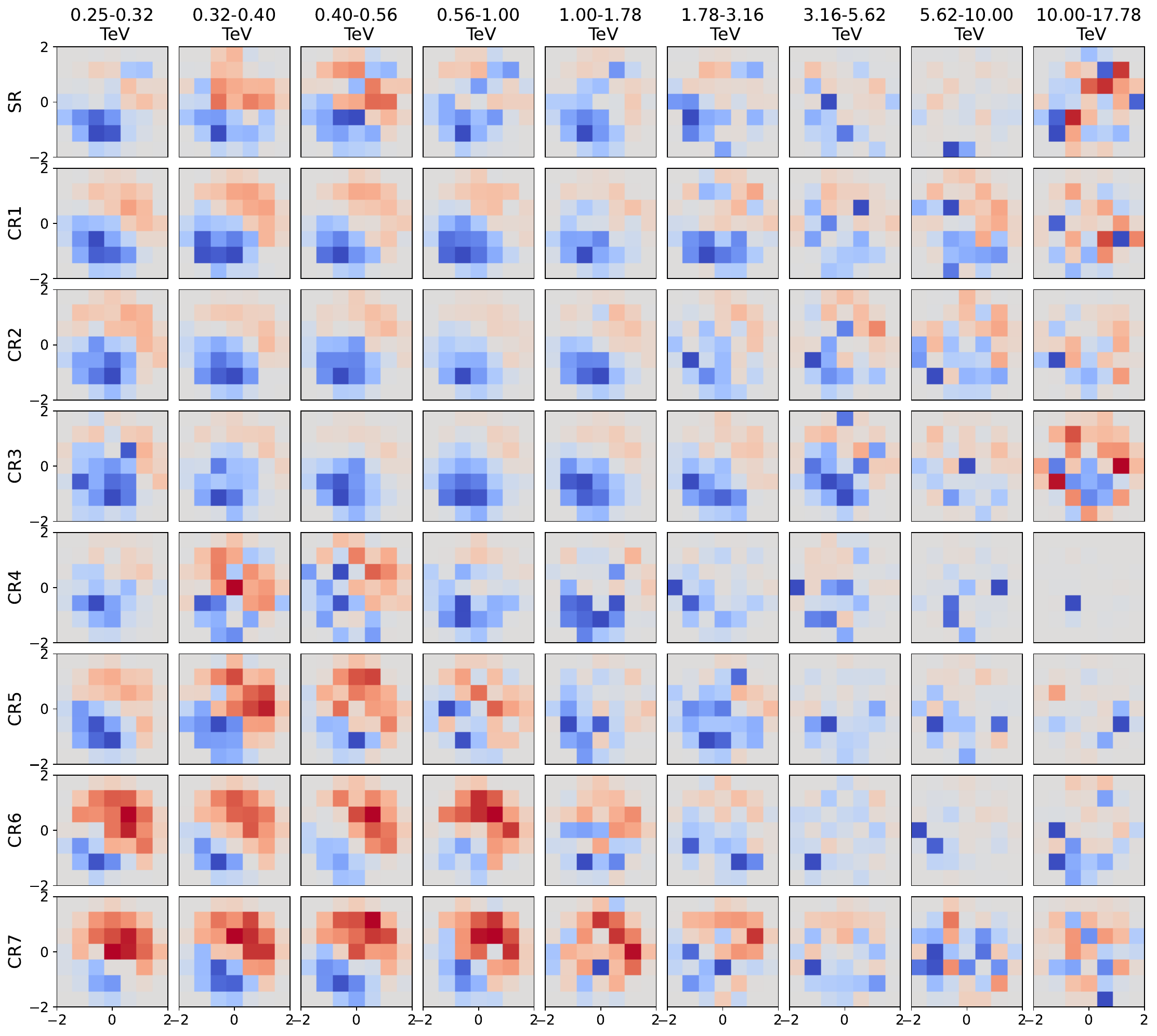}
\caption{
Exemplary singular vectors $\vec{v}_{k}$ with $k=0$ (top) and $k=5$ (bottom), constructed from the distributions of cosmic-ray events at the reconstructed energies.
The lower-order singular vector gives the generic shape of the background function, while the higher-order vector gives more detailed features. 
The color scales are centered at zero but are not the same across different mini-maps; they are saturated for the bin containing the largest absolute amplitude in a mini-map.
}
\label{fig:eigenvectors}
\end{figure*}

The relevance of a singular vector $\vec{v}_{k}$ can be evaluated by the corresponding singular value $\sigma_{k}$.
For example, Figure \ref{fig:singular_values} shows the singular values of the template matrix $M$ as a function of $k$.
The singular value $\sigma_{k}$ decreases rapidly as $k$ increases until $k \sim 10$.
The singular vectors with $k\leq 10$ are the significant vectors that are relevant to $M$ and will be most-significant to model the background component of the ON data profiles.
For example, for the energy range $E \in [0.25,17.8]$ TeV, Figure \ref{fig:eigenvectors} shows the singular vectors for $k=0$ and $k=5$.
While the first singular vector $\vec{v}_{0}$ (top panel) shows a generic feature of the cosmic-ray distribution,
the higher-order singular vector $\vec{v}_{5}$ (bottom panel) captures more variations, such as the gradient along the vertical axis at low energies ($[0.25,1.0]$ TeV), which are sensitive to elevation effects like the depth of atmosphere and the detector effective area.

\subsection{Construction of background models with singular vectors}

The STOICS method approximates the true background profile for the ON data as a linear combination of the singular vectors, i.e.
\begin{equation}
\vec{\lambda} = 
\sum_{k=1}^{k_{c}} a_{k} \vec{v}_{k},
\label{eq:singular_vector_model}
\end{equation}
where $\vec{\lambda}$ is the background model, $a_{k}$ is the weight associated with the singular vector $\vec{v}_{k}$, and $k_{c}$ is the number of singular vectors to be included.

We define a pixel-wise Poisson likelihood function as
\begin{equation}
L_{j}(x_{j}|\lambda_{j}) 
=
\frac{\lambda_{j}^{x_{j}} e^{-\lambda_{j}}}{x_{j}!},
\end{equation}
where $\lambda_{j}$ is the predicted background count in the $j$-th pixel of the profile and $x_{j}$ is the observed count.

The STOICS method finds the solution for the coefficients $a_{k}$ by maximizing the likelihood sum in the cosmic-ray-like regions
\begin{equation}
\frac{\partial}{\partial \vec{a}}
\left(
\sum_{j \in \mathrm{CR}} \log L_{j}(\vec{a})
\right) = 0.
\end{equation}
The pixels in the $\gamma$-like regions (SR) are kept blind in the likelihood maximization process.
In the maximization process, runs are grouped by zenith angle into small batches (each batch with exposure of $\sim2$-hours), and ON data within a batch are summed to reduce statistical fluctuations.

The background profile $\vec{\lambda}$ represents the background distribution in the camera frame.
To make background prediction in the sky-coordinate system, the background distribution is constructed from events in the CR-like regions normalized with a weight
\begin{equation}
  w_{ij} =
  \begin{cases}
    \lambda_{i}/x_{j}, & \text{if } j \text{ corresponds to } i \\
    0, & \text{else}
  \end{cases}
\end{equation}
where $i$ is a pixel in the $\gamma$-like region and $j$ is the corresponding pixel in the camera frame in the CR-like region of the same energy interval. The background count at the sky coordinate $\vec{r}$ is then calculated as
\begin{equation}
N_{\mathrm{bkg},i}(\vec{r}) = \sum_{j\in \mathrm{CR}} w_{ij} N_{\mathrm{CR},j}(\vec{r}).
\end{equation}
The construction of a background model from the reweighted CR-like events has been discussed in \cite{Rowell_Template2003}, which shows the ability to compensate for localized systematic changes such as the presence of stars in the field of view.

It should be noted that if the source morphology is known, one could allow $\gamma$-ray-free pixels in $\gamma$-like regions to participate in likelihood maximization to improve the accuracy of background modeling.

\subsection{Model complexity: $k_{\mathrm{c}}$ and pixel size}

The complexity of the STOICS method is controlled by the number of singular vectors $k_{c}$ and the size of the pixel area $A_{\mathrm{pix}}$ in the template matrix $M$.
A smaller $A_{\mathrm{pix}}$ allows the method to model the details of the angular variations of the background rate.
A larger $k_{\mathrm{c}}$ gives the method higher degrees of freedom to model the variations across observations taken under different conditions.
The size of the pixels should be chosen
to balance between having sufficient statistics in each pixel and maintaining
the shape of the cosmic-ray distribution.
Similarly, higher-order singular vectors also suffer from statistical fluctuations, and $k_{\mathrm{c}}$ should be chosen accordingly to avoid noise-dominated singular vectors.

\begin{figure*}
\centering
\includegraphics[width=0.30\linewidth]{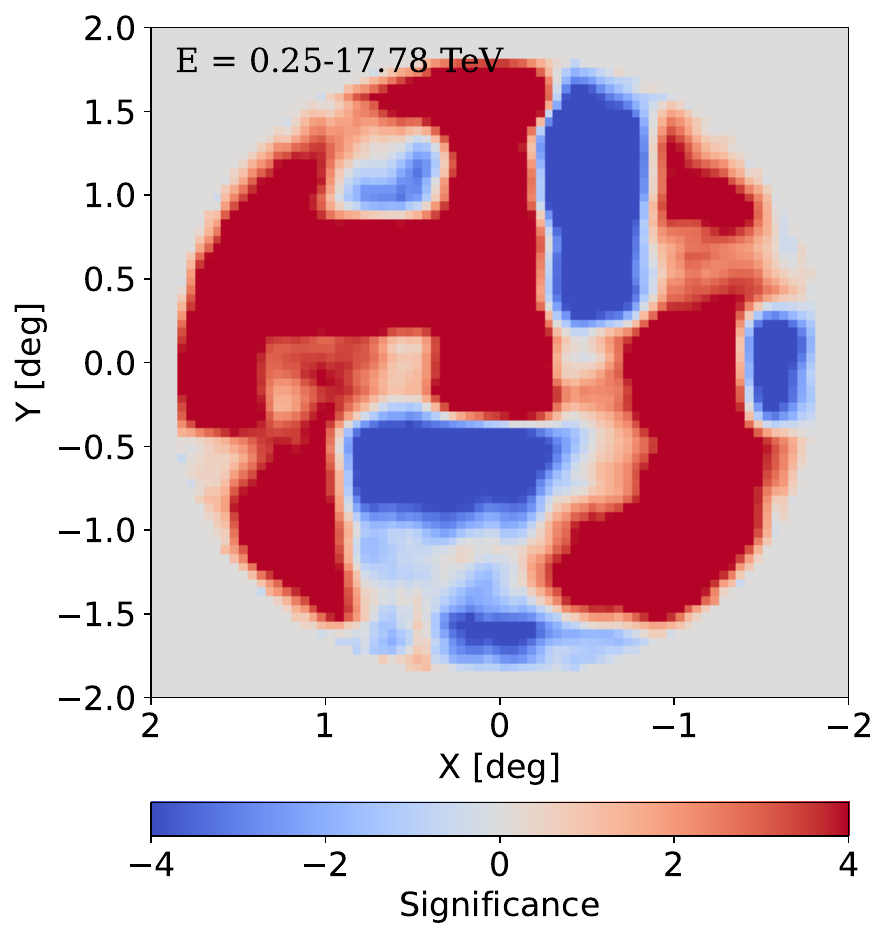}
\includegraphics[width=0.45\linewidth]{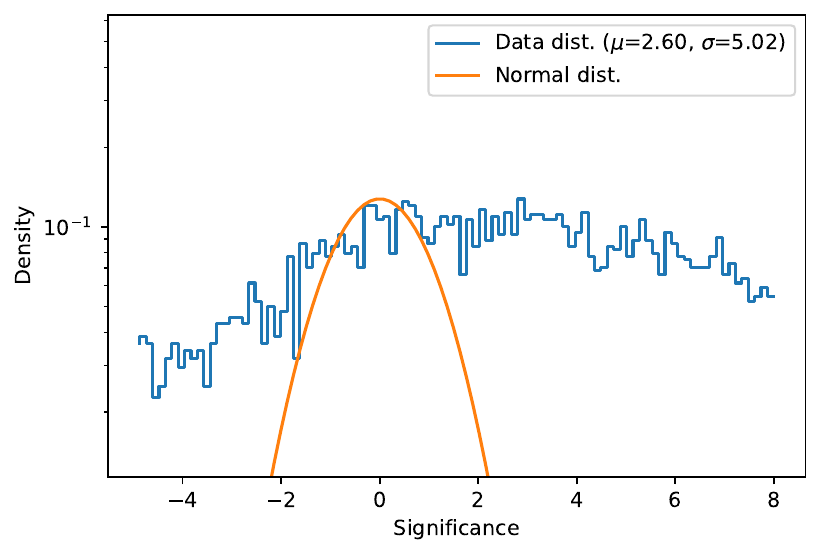}\\
\includegraphics[width=0.30\linewidth]{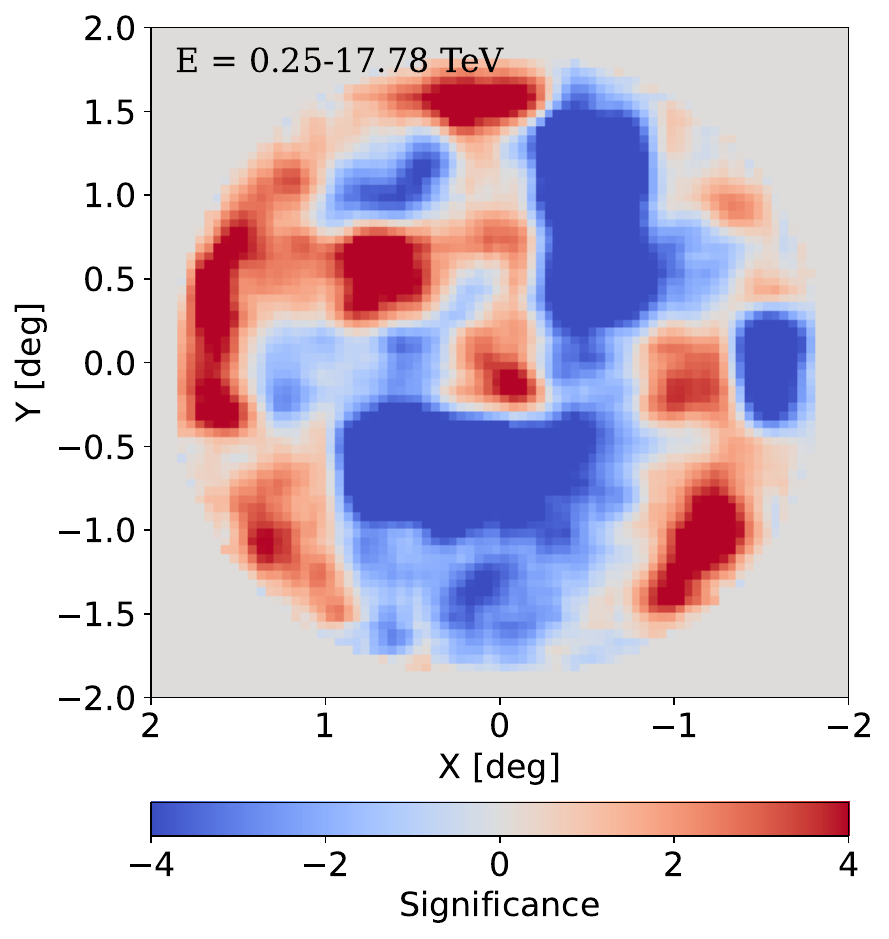}
\includegraphics[width=0.45\linewidth]{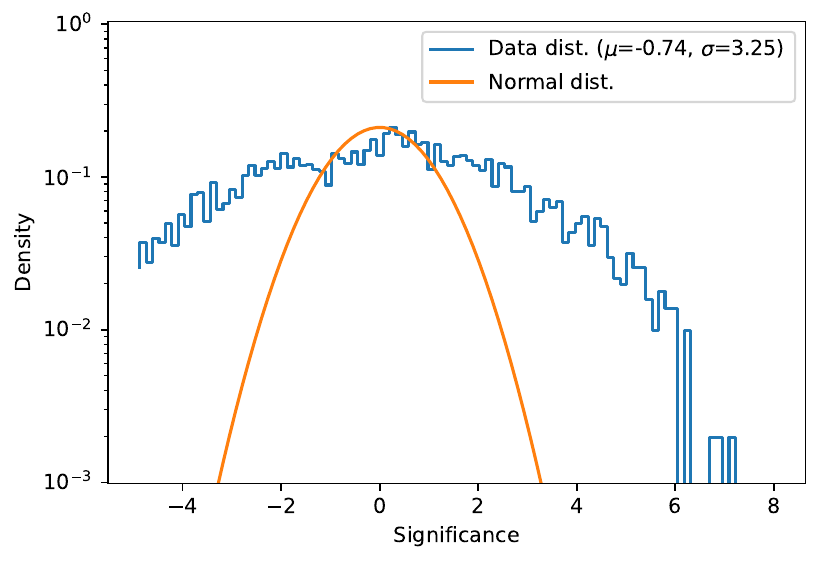}\\
\includegraphics[width=0.30\linewidth]{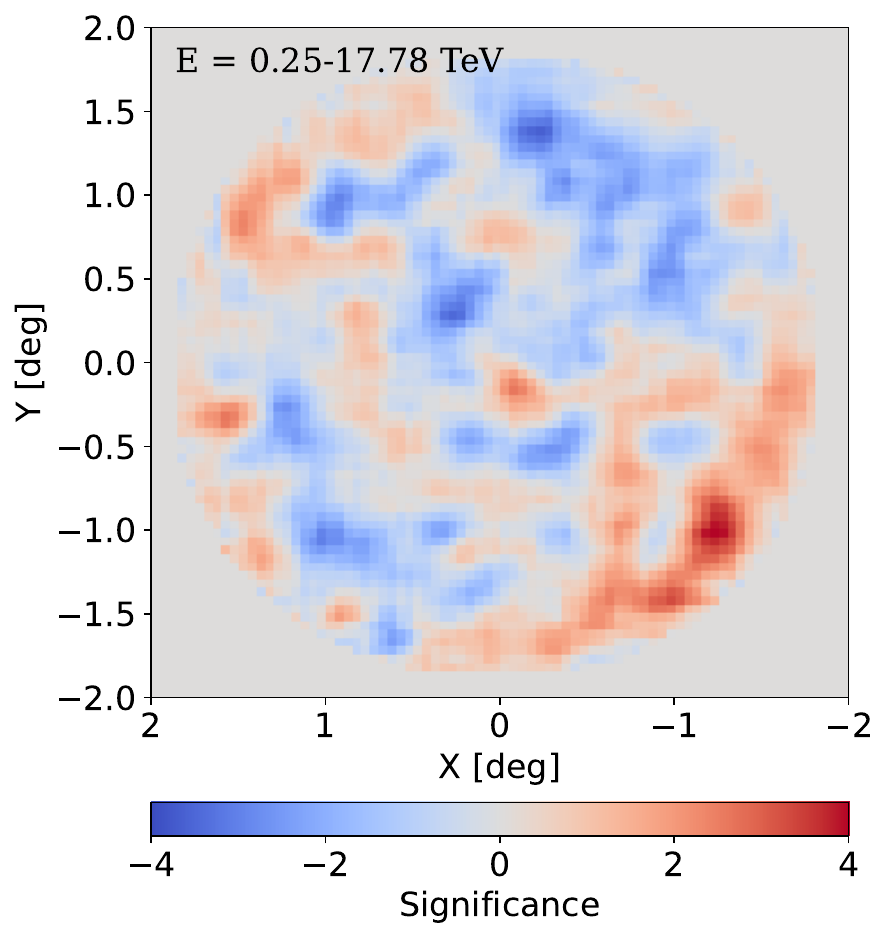}
\includegraphics[width=0.45\linewidth]{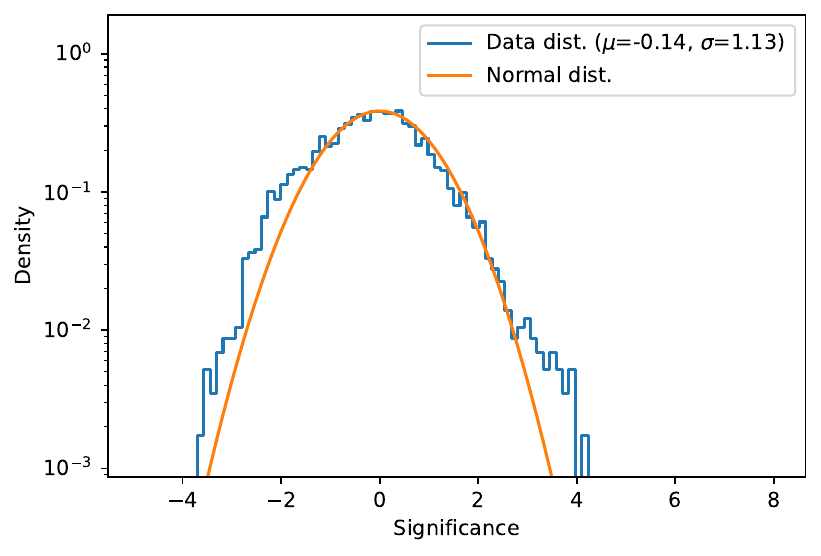}\\
\includegraphics[width=0.30\linewidth]{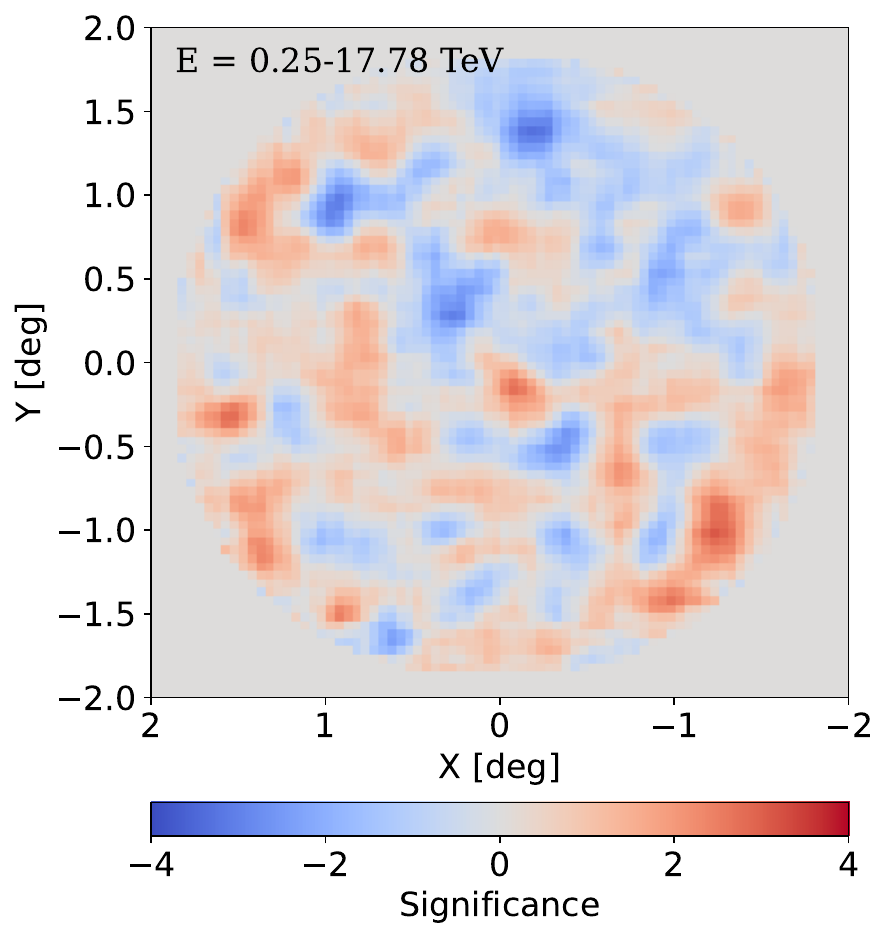}
\includegraphics[width=0.45\linewidth]{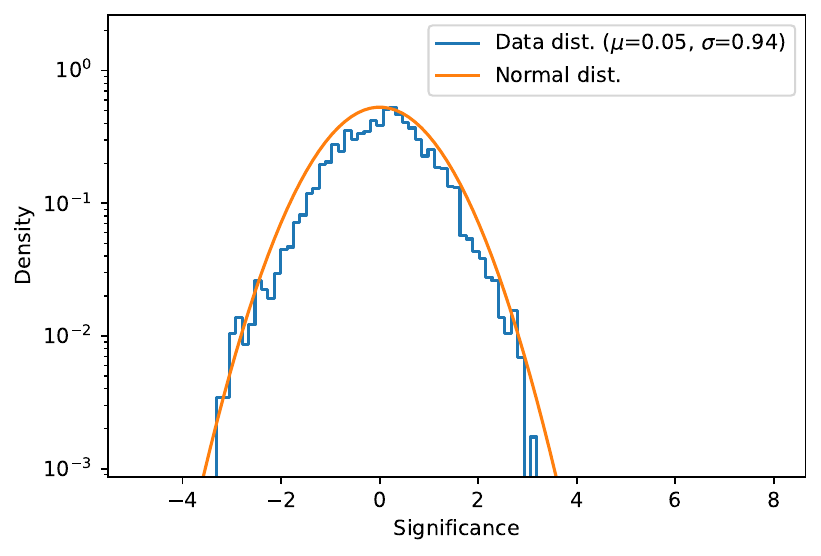}
\caption{
Exemplary maps of model error significance in camera frame and the significance distributions (mean $\mu$ and RMS $\sigma$) in energy range $E \in [0.25,17.8]$ TeV. 
The background models are using $k_{\mathrm{c}}=1$ (top-left), $k_{\mathrm{c}}=2$ (top-right), $k_{\mathrm{c}}=4$ (bottom-left), and $k_{\mathrm{c}}=32$ (bottom-right).
The figure demonstrates that by including more singular vectors (higher $k_{\mathrm{c}}$), the background model achieves better accuracy.
}
\label{fig:systematics_eigenvector_LE}
\end{figure*}

Figure \ref{fig:systematics_eigenvector_LE} shows the significance maps of the modeling error for different $k_{\mathrm{c}}$ and the significance distributions using blank sky observations of a total of 135-hour exposure containing no detected $\gamma$-ray sources.
With more singular vectors included in the modeling, it is observed that the background model becomes more accurate.
The choice of the number of singular vectors in the model depends on the exposure of the observations.
We found that the background model should optimally include at least 32 singular vectors for observations with exposure $>10^{5}\ \mathrm{m}^{2} \times 100\ \mathrm{hr}$.
The effect of the number of singular vectors in the model is discussed in more detail in Section \ref{sec:syst_unc}.

\begin{figure*}
\centering
\includegraphics[width=0.30\linewidth]{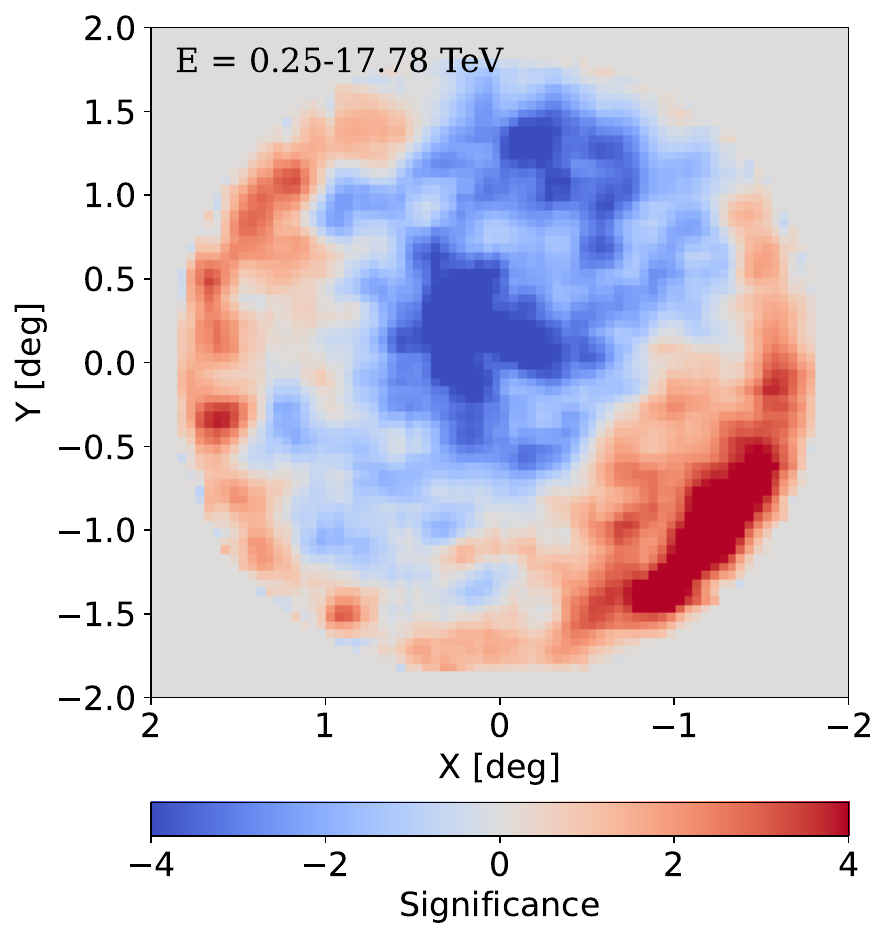}
\includegraphics[width=0.45\linewidth]{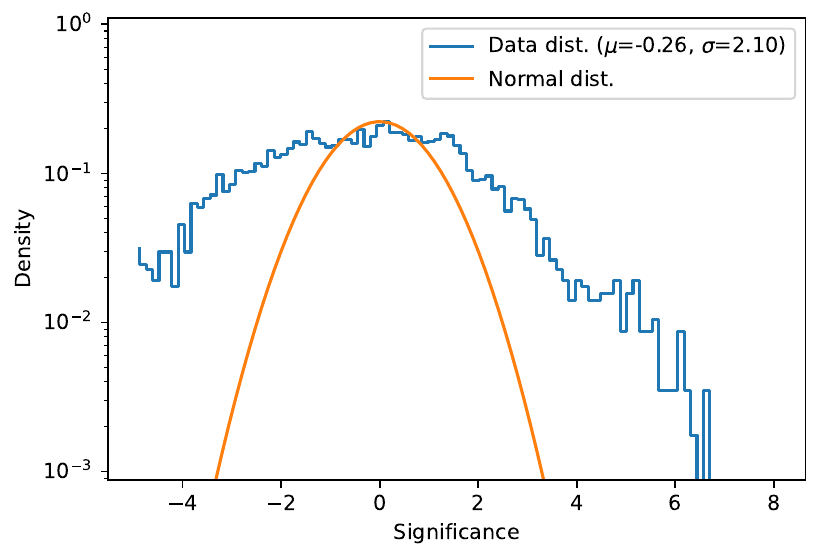}\\
\includegraphics[width=0.30\linewidth]{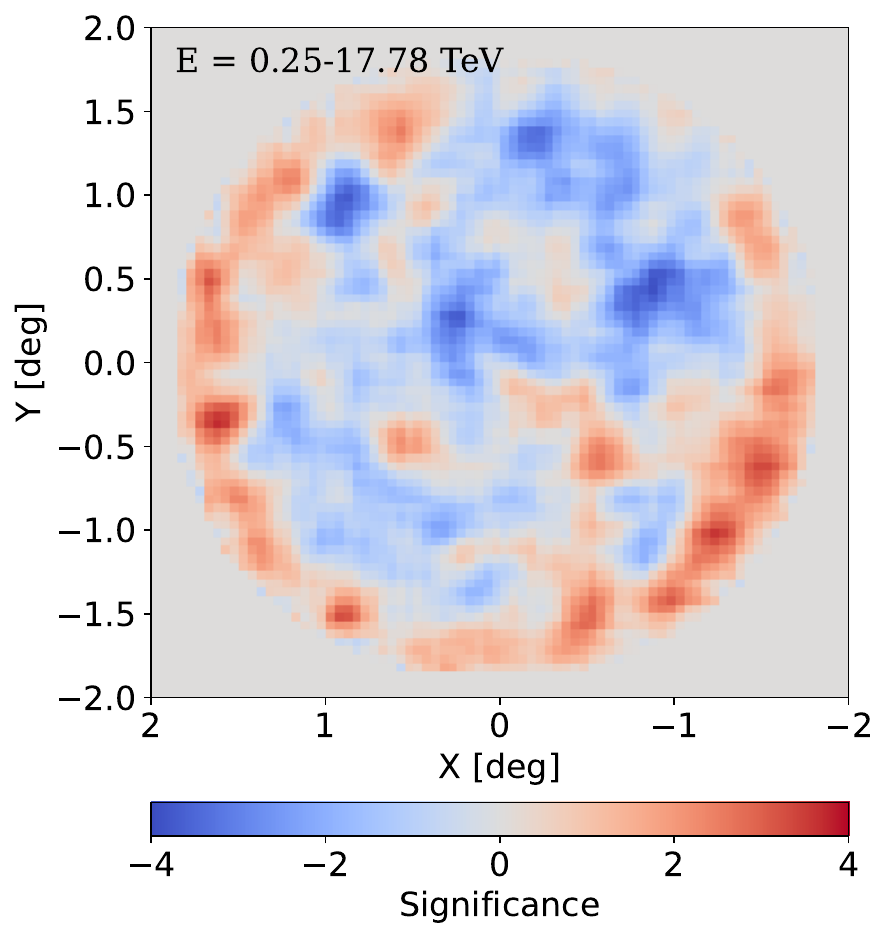}
\includegraphics[width=0.45\linewidth]{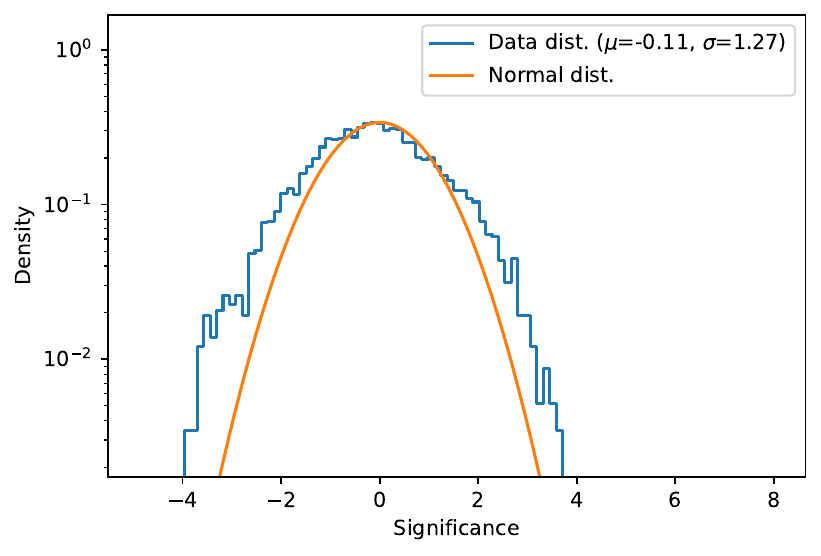}\\
\includegraphics[width=0.30\linewidth]{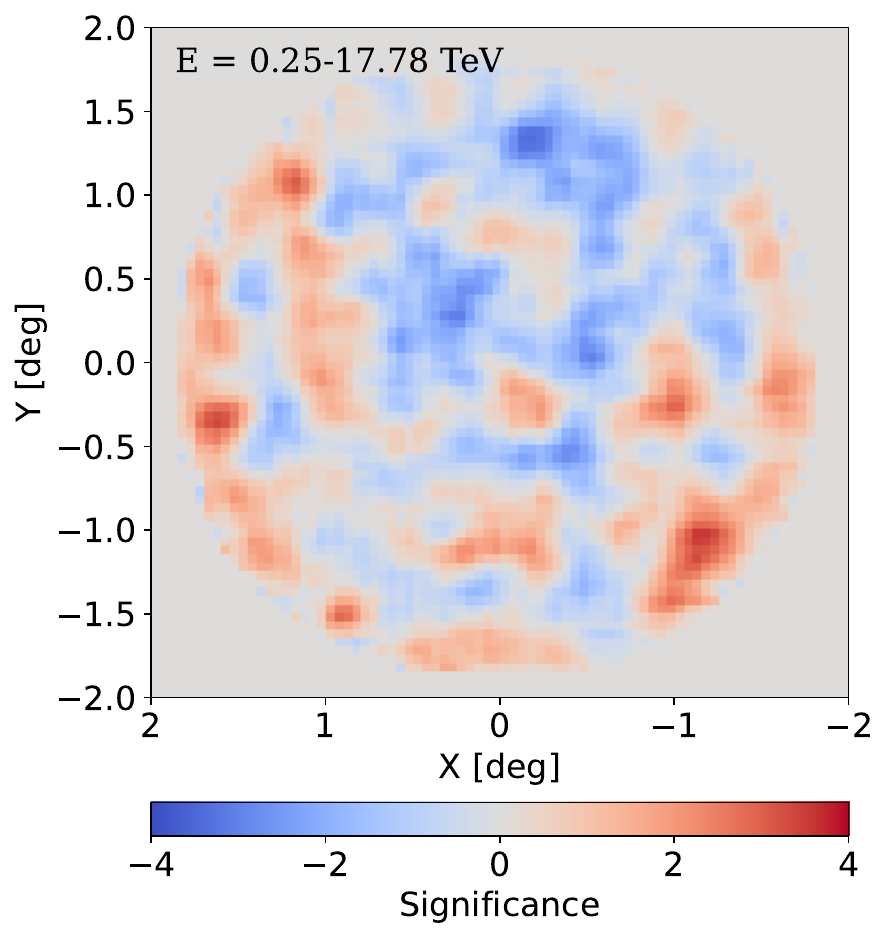}
\includegraphics[width=0.45\linewidth]{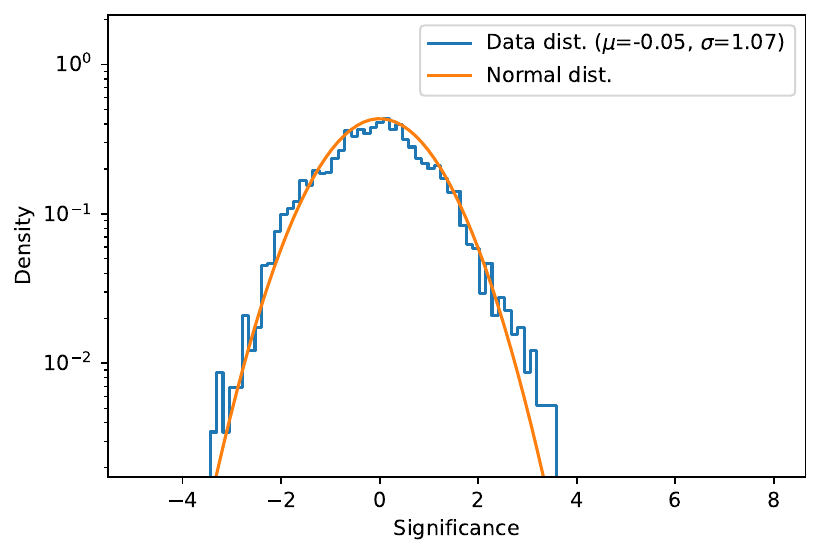}\\
\includegraphics[width=0.30\linewidth]{significance_coolwarm_sky_map_NGC1275_allE_fullspec32.pdf}
\includegraphics[width=0.45\linewidth]{significance_distribution_NGC1275_allE_fullspec32.pdf}\\
\caption{
Exemplary maps of model error significance in camera frame and the significance distributions (mean $\mu$ and RMS $\sigma$) in energy range $E \in [0.25,17.8]$ TeV. 
The background models are using pixel sizes of $3.4^{\circ} \times 3.4^{\circ}$ (top-left), $1.3^{\circ} \times 1.3^{\circ}$ (top-right), $0.8^{\circ} \times 0.8^{\circ}$ (bottom-left), and $0.57^{\circ} \times 0.57^{\circ}$ (bottom-right).
The figure demonstrates that the background model achieves better accuracy with finer pixel sizes.
}
\label{fig:systematics_pixel_size}
\end{figure*}

In addition to the number of singular vectors, we find that the modeling accuracy is also sensitive to the choice of pixel size.
Figure \ref{fig:systematics_pixel_size} shows the significance maps of the modeling error for different pixel sizes and the significance distributions using blank sky observations that do not contain $\gamma$-ray sources.
In the exaggerated case of the entire camera FoV as a single pixel (top left, $3.4^{\circ} \times 3.4^{\circ}$), we observe significance errors in background modeling with an overestimated background rate in the central region of the field of view and an underestimated background rate around the edge.
The background model improves as we reduce the size of pixels. We found that the pixel size should be smaller than $0.57^{\circ} \times 0.57^{\circ}$ (bottom-right plot)  for observations with exposure $>10^{5}\ \mathrm{m}^{2} \times 100\ \mathrm{hr}$.

\section{Estimation of systematic uncertainty}
\label{sec:syst_unc}

For a template matrix $M$ containing $m$ OFF runs and $n$ pixels,
\begin{equation}
M = \sum_{k=1}^{r} \sigma_{k} \vec{u}_{k}\vec{v}_{k}^{\top},
\end{equation}
where $r=\mathrm{min}(m,n)$ is the rank of $M$.
We truncate the rank to $k=k_{c}$ to avoid singular vectors dominated by noise and approximate the background model using Equation \ref{eq:singular_vector_model}.
An important question to ask is what the size of the systematic uncertainty is due to the truncation of singular vectors and the choice of the OFF runs.

\begin{figure*}
\centering
\includegraphics[width=0.24\linewidth]{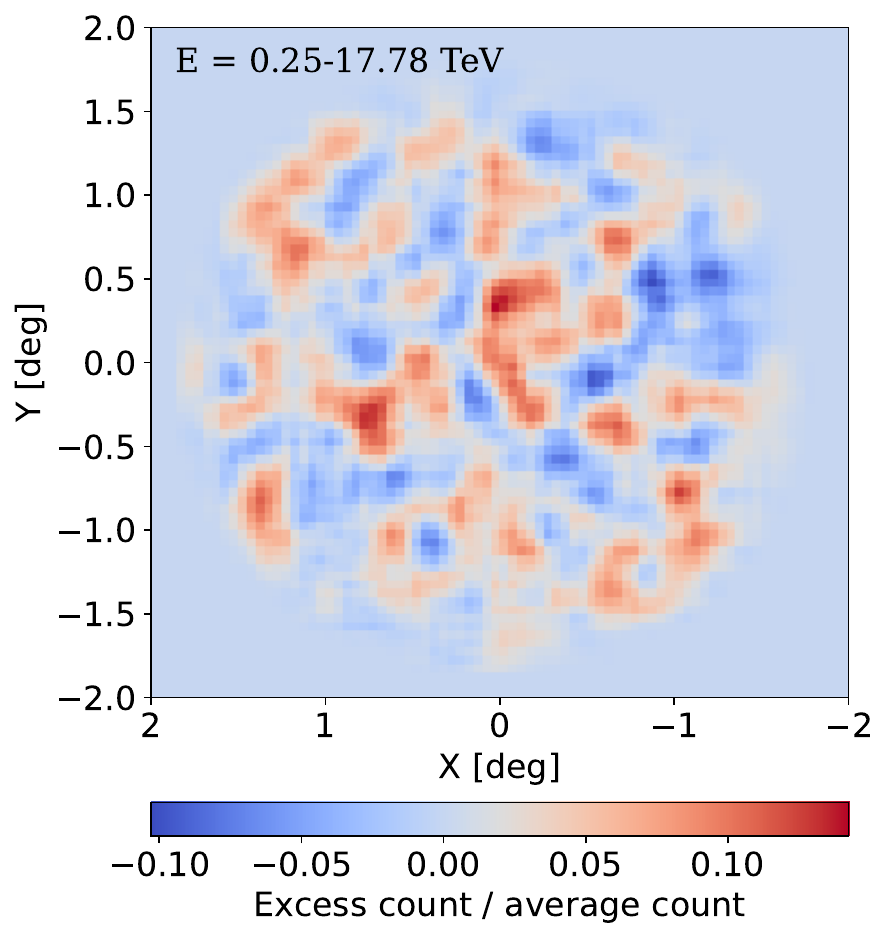}
\includegraphics[width=0.24\linewidth]{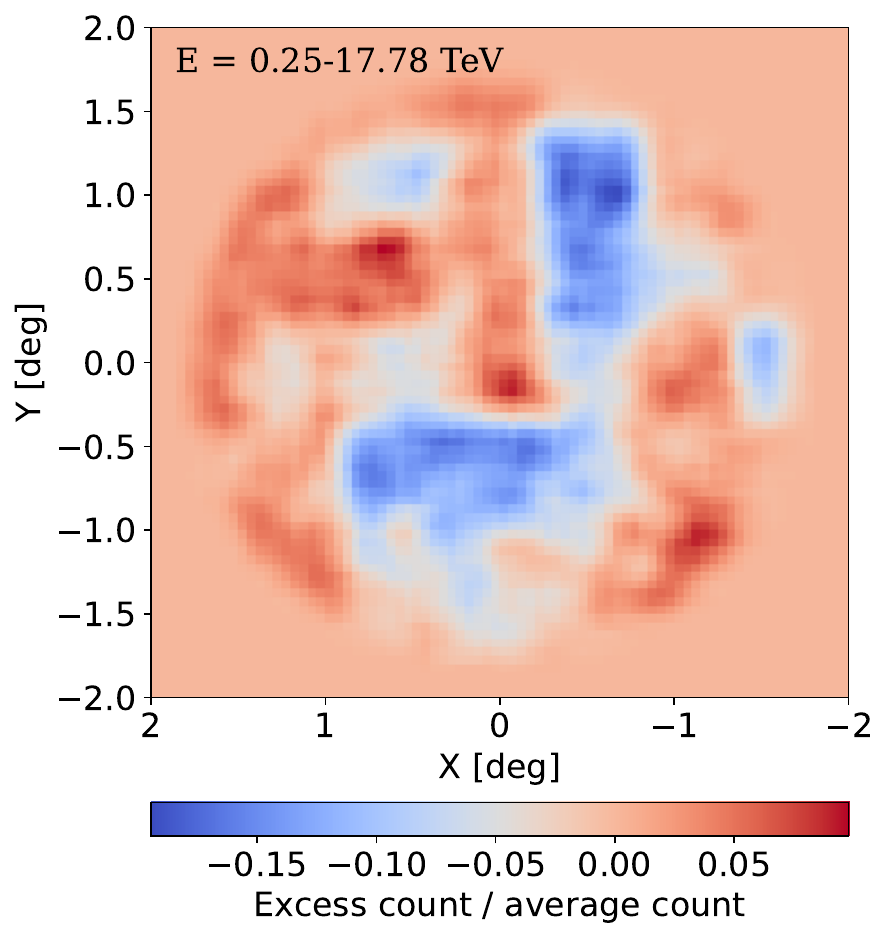}
\includegraphics[width=0.24\linewidth]{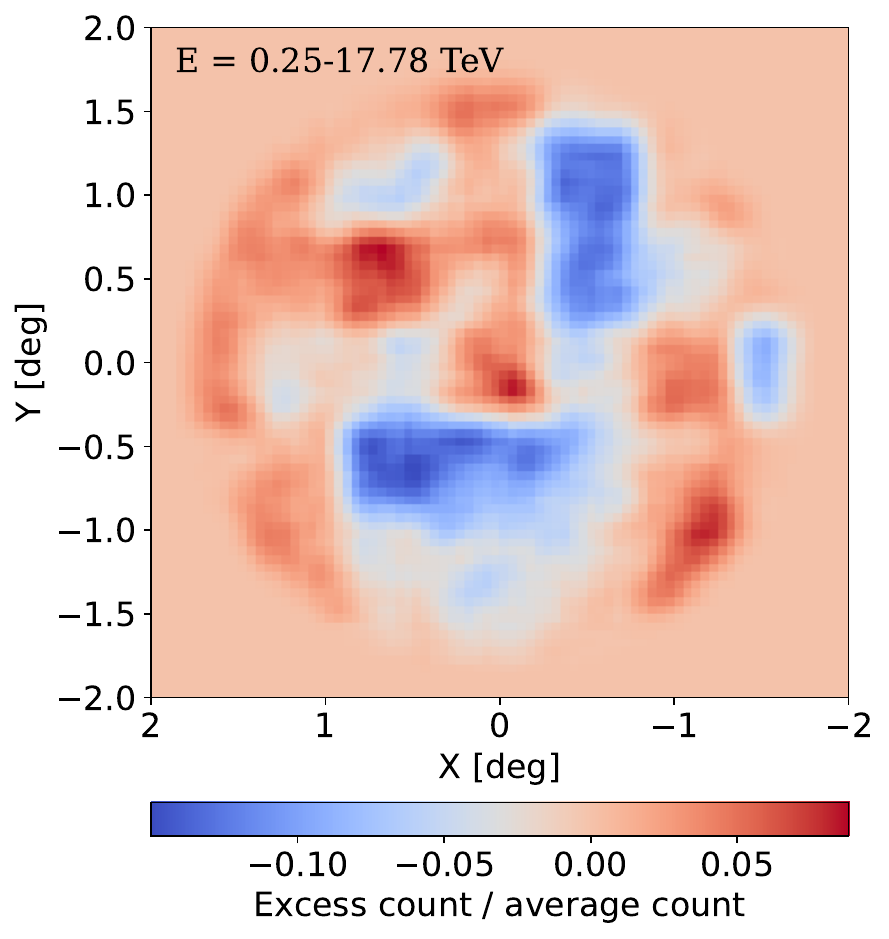}
\includegraphics[width=0.24\linewidth]{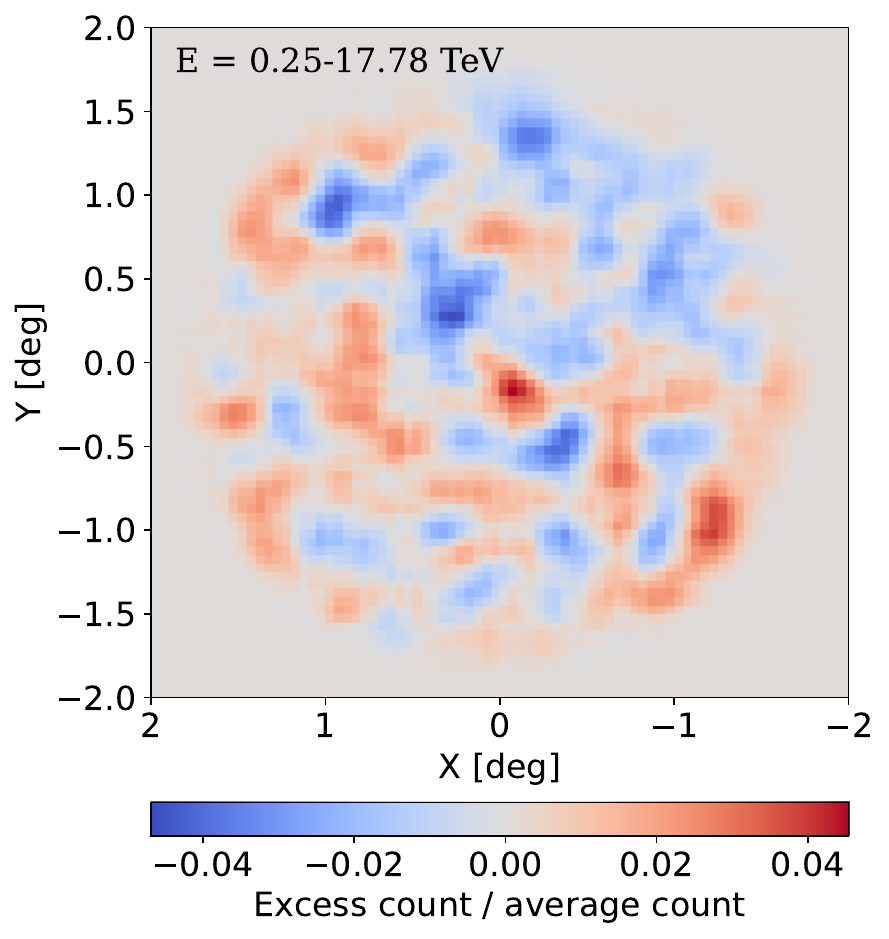}
\\
\includegraphics[width=0.24\linewidth]{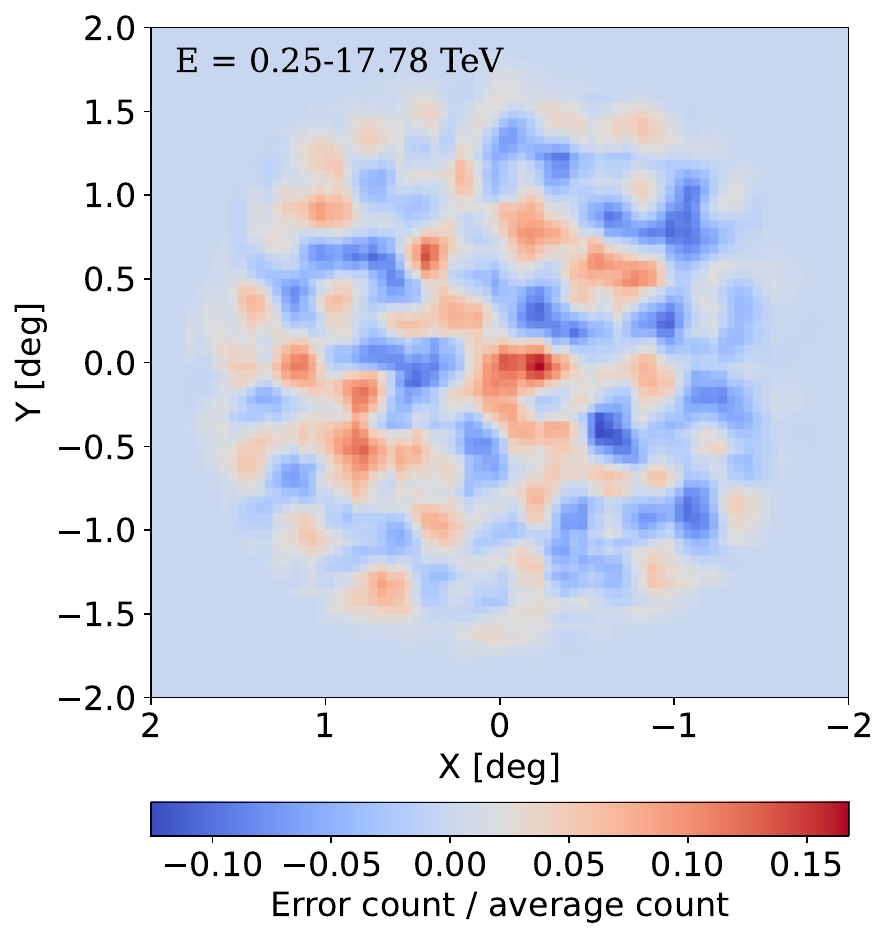}
\includegraphics[width=0.24\linewidth]{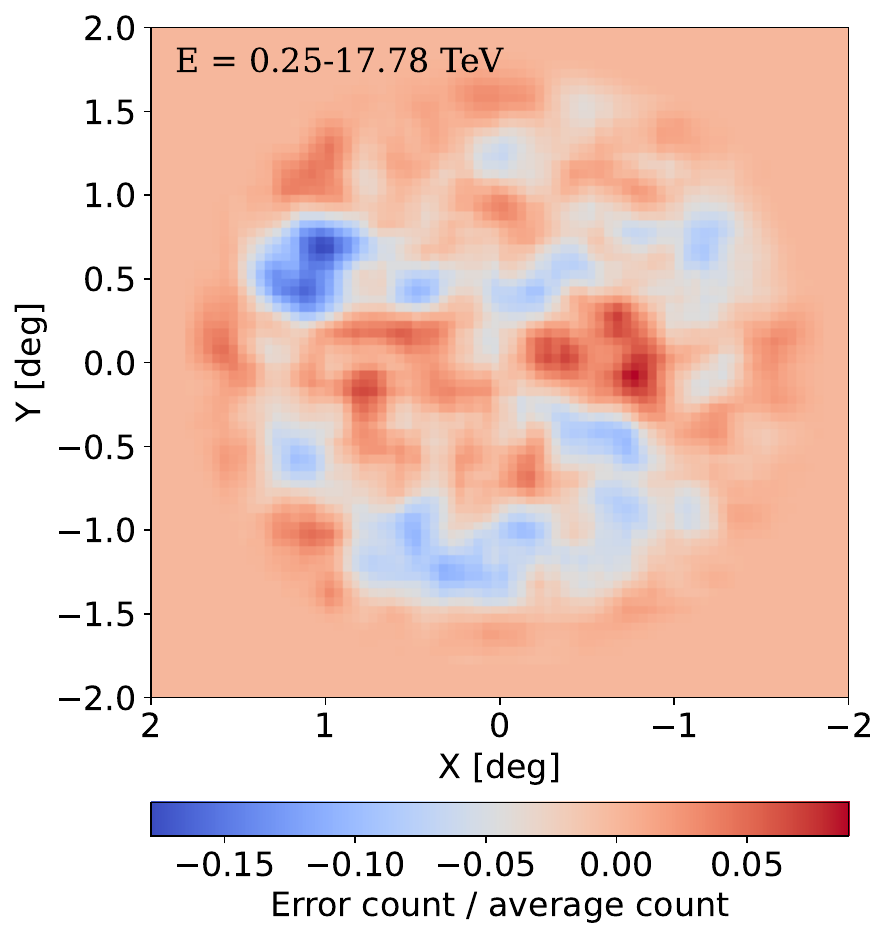}
\includegraphics[width=0.24\linewidth]{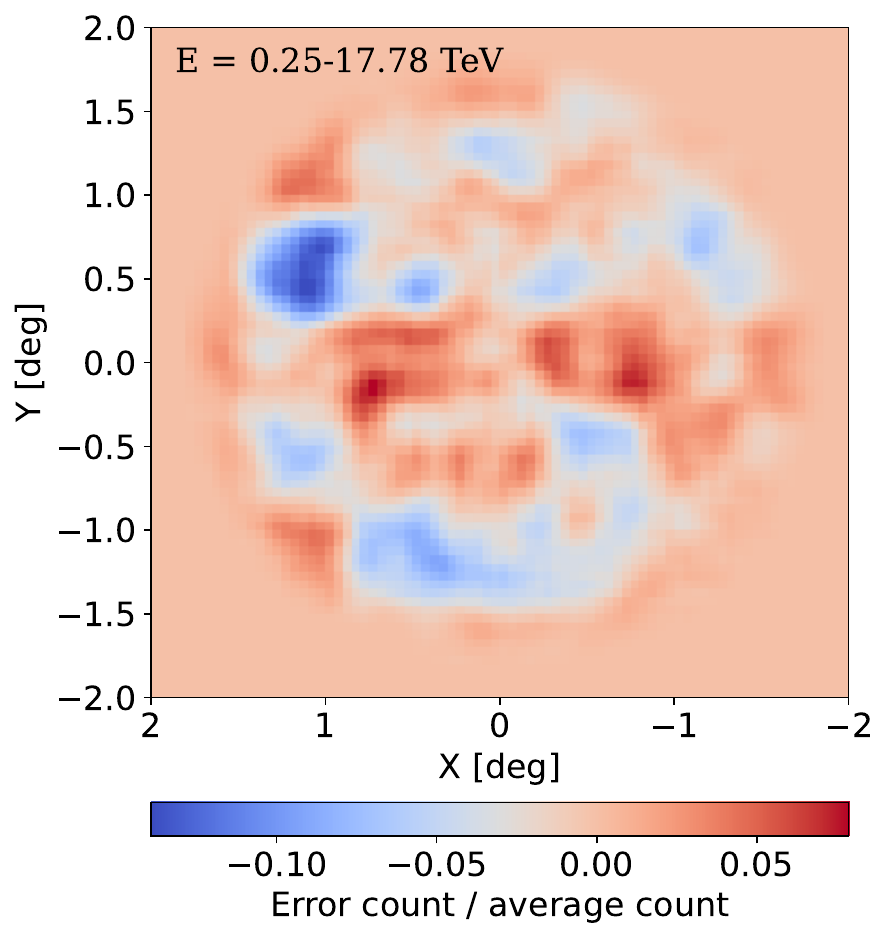}
\includegraphics[width=0.24\linewidth]{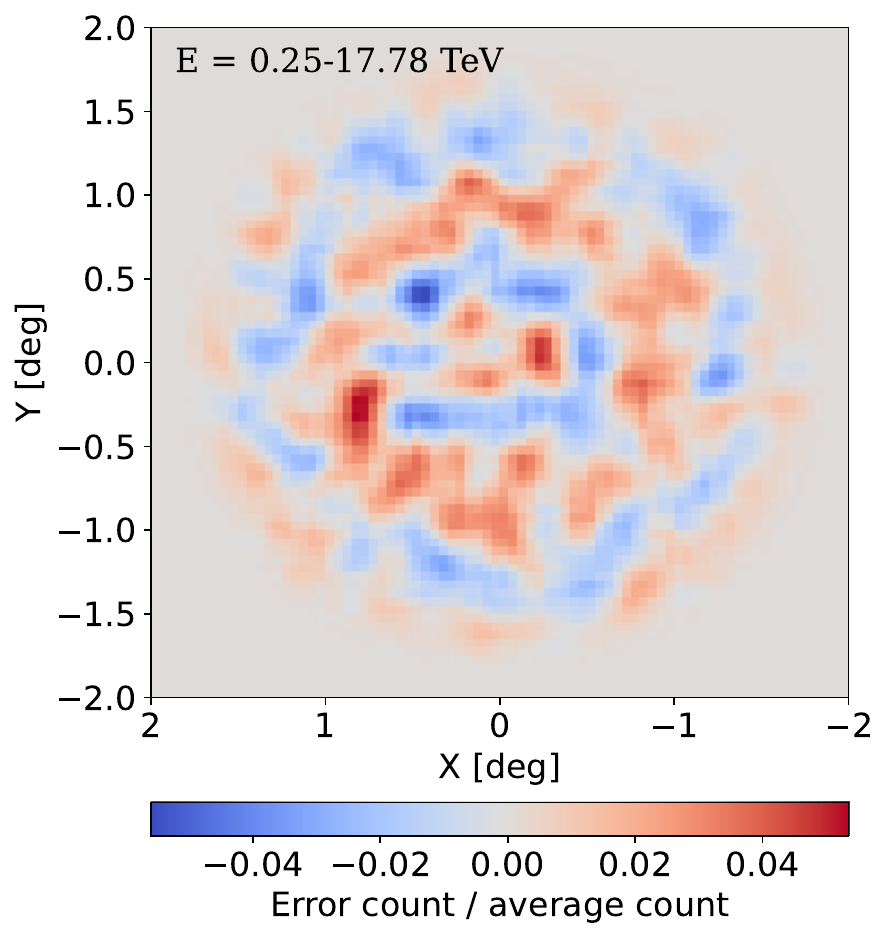}
\caption{
The correlation between the model accuracy in the $\gamma$-like region and the accuracy in the CR-like region is demonstrated by the residual maps in the $\gamma$-like region (top row) and in the CR-like region (CR1, bottom row) for the analyses of various exposures.
The first column shows the analysis using $k_{c}=2$ for 10-hours data, the second column shows the analysis using $k_{c}=2$ for 50-hours data, the third column shows the analysis using $k_{c}=2$ for 135-hours data, and the fourth column shows the analysis using $k_{c}=32$ for 135-hours data.
These plots demonstrate that while for short observation periods two singular vectors are sufficient to generate a background model that does not show clear systematic mis-modeling, for longer exposures, the background is modeled inaccurately (as evidenced from the non-uniform maps), and a larger number of singular vectors is necessary for the 135-hours dataset.
}
\label{fig:systematic_estimation}
\end{figure*}

The size of the systematic uncertainty $\theta_{\mathrm{syst}}$ can be estimated by observing the residuals in a chosen CR-like region,
\begin{equation}
\theta^{2}_{\mathrm{syst}} = \frac{\lambda_{\mathrm{SR}}^{2}}{\lambda_{\mathrm{CR}}^{2}}
\left(x_{\mathrm{CR}}-\lambda_{\mathrm{CR}}\right)^{2},
\end{equation}
where $\lambda_{\mathrm{SR}}$ is the background prediction in a pixel in SR, $\lambda_{\mathrm{CR}}$ is the background prediction in the corresponding pixel in CR, and $x_{\mathrm{CR}}$ are the observed data in the pixel in CR.
Figure \ref{fig:systematic_estimation} demonstrates the systematic uncertainty estimation using the residuals in CR1.
The first column of Figure \ref{fig:systematic_estimation} shows the residual map of an analysis using $k_{c}=2$ for a 10-hour data in the $\gamma$-like region (top panel) and in the CR-1 region (bottom panel). 
We observe no significant systematic mis-modeling in either region.
As we increase the exposure to 50 hours in the second column, the systematic error surpasses the statistical fluctuation, and the systematic structure emerges in both the $\gamma$-like region (top panel) and the CR-like region (bottom panel).
The systematic artifacts become more significant as exposure increases to 135 hours in the third column.
In the fourth column, the analysis increases the number of singular vectors to $k_{c}=32$, and the systematic mis-modeling is reduced in both $\gamma$-like and CR-like regions.
The observations provided in Figure \ref{fig:systematic_estimation} show that the accuracy of the background model in the $\gamma$-like region is correlated with the quality of fit in the CR-like region, and therefore the CR-like region is a good indicator of the systematic uncertainty of the background model.

To further study the correlation between the model accuracy in the $\gamma$-like region and the accuracy in the CR-like region, we analyzed a total of 3158 hours of data collected by VERITAS.
These data are composed of extragalactic observations of point-like sources (AGNs and dwarf galaxies), and events with arrival directions pointing back to known sources are vetoed to ensure that the data are $\gamma$-ray-free.
These observations are grouped into several batches so that each batch has an average exposure of $100$ hours. 
We define a metric of error
\begin{equation}
\epsilon_{\mathrm{SR/CR}} = \frac{1}{\sum_{i} T_{i}}
\sum_{i} T_{i}\frac{ \|x_{\mathrm{SR/CR},i} - \lambda_{\mathrm{SR/CR},i}\|}{x_{\mathrm{SR/CR},i}},
\end{equation}
where $\lambda_{\mathrm{SR/CR},i}=\sum_{j \in \mathrm{SR/CR}} \lambda_{j}$ is the integrated predicted background count in the SR (or the CR) of the $i$-th batch, $x_{\mathrm{SR/CR},i}=\sum_{j \in \mathrm{SR/CR}} x_{j}$ is the integrated observed count and $T_{i}$ is the exposure time of the observation batch.

\begin{figure*}
\centering
\includegraphics[width=0.48\linewidth]{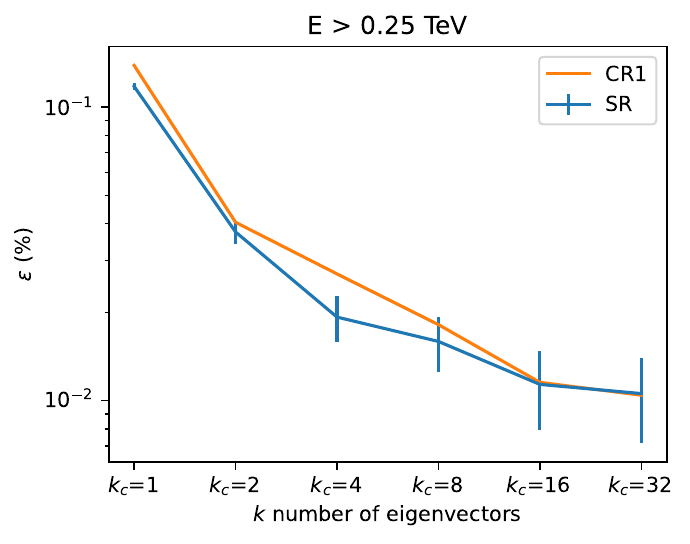}
\includegraphics[width=0.48\linewidth]{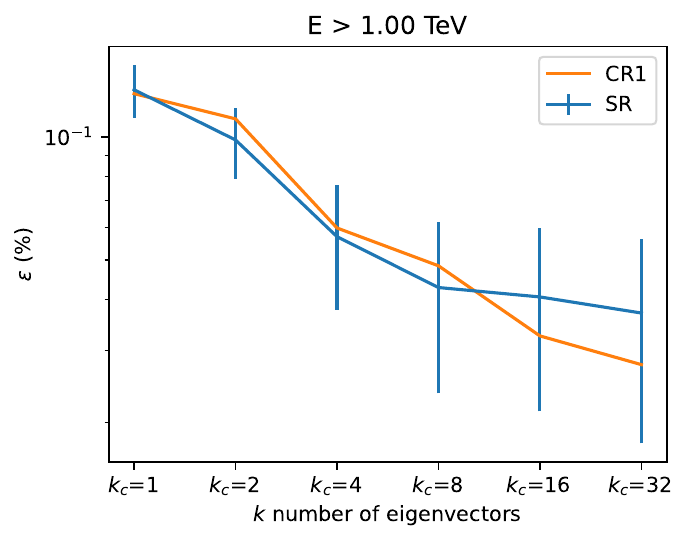}
\caption{
The error $\epsilon$ measured as a function of $k_{c}$ in the $\gamma$-like region (blue curve) and in the CR1 region (orange curve) in two energy ranges. 
The vertical bars represent the averaged statistical uncertainty of $x_{\mathrm{SR}}$.
The modeling error in the $\gamma$-like region and the error in the CR region show strong correlation.
}
\label{fig:kc_vs_error}
\end{figure*}

The error $\epsilon_{\mathrm{SR/CR}}$ is measured as a function of $k_{c}$ and is shown in Figure \ref{fig:kc_vs_error} for both $\gamma$-like and CR-like regions.
At low energy ($E<1$ TeV), $\epsilon_{\mathrm{SR}}$ decreases quickly for the increase of $k_{c}$ until $k_{c}=16$.
The difference between $\epsilon_{\mathrm{SR}}(k_{c}=16)$ and $\epsilon_{\mathrm{SR}}(k_{c}=32)$ is smaller than the statistical fluctuation (indicated by the vertical bar), and this means that $\epsilon(k_{c}=32)$ is sufficient for an analysis of the $\sim 100$-hour data.
At high energies ($E>1$ TeV), $\epsilon_{\mathrm{SR}}$ is insensitive to the change of $k_{c}$ for $k_{c}>4$ because the singular vectors are dominated by statistical noise, and only $\sim 4$ singular vectors are useful for constructing the background model.
The orange curves in Figure \ref{fig:kc_vs_error} show the $\epsilon_{\mathrm{CR}}$ using the residuals in CR1, and we observe that the size of $\epsilon_{\mathrm{SR}}$ is nicely predicted by $\epsilon_{\mathrm{CR}}$.

The systematic uncertainty will inflate the estimated significance, and thus we use the estimated systematic uncertainty to update the detection significance.
We first compute the statistical significance as
\begin{equation}
S_{\mathrm{stat}} =
\sqrt{
-2 \ln \frac{\lambda_{\mathrm{SR}}^{x_{\mathrm{SR}}} e^{-\lambda_{\mathrm{SR}}}}{x_{\mathrm{SR}}!}
},
\label{eq:stat_error}
\end{equation}
then we define the systematic significance as
\begin{equation}
S_{\mathrm{syst}} = \frac{x_{\mathrm{SR}}-\lambda_{\mathrm{SR}}}{\theta_{\mathrm{syst}}},
\label{eq:syst_error}
\end{equation}
and finally, the updated significance is
\begin{equation}
S_{\mathrm{total}}^{2} = \left( \frac{1}{S_{\mathrm{stat}}^{2}}+\frac{1}{S_{\mathrm{syst}}^{2}} \right)^{-1}.
\label{eq:updated_significance}
\end{equation}
In this way, we ensure that the a detection with significance $\geq 5 \sigma$ is not inflated by the systematic uncertainty.
We also note that Equation \ref{eq:stat_error} neglects the statistical errors in the singular templates and Equation \ref{eq:syst_error} neglects the statistical error in CR1, so the significance is slightly underestimated, leading to slightly narrower significance distributions demonstrated in this paper.

\section{Examples}
\label{sec:example}

The STOICS uses loose selections for the matched OFF runs and construct `singular' background templates by performing a principal component analysis of the matched runs.
These loosely selected OFF runs cover a wide range of observing conditions, allowing the singular templates to model subtle variations in the background distributions.
In this section, we show the background modeling for observations of various source types.

\subsection{Extragalactic point-like sources}

\begin{figure*}
\centering
\includegraphics[width=0.32\linewidth]{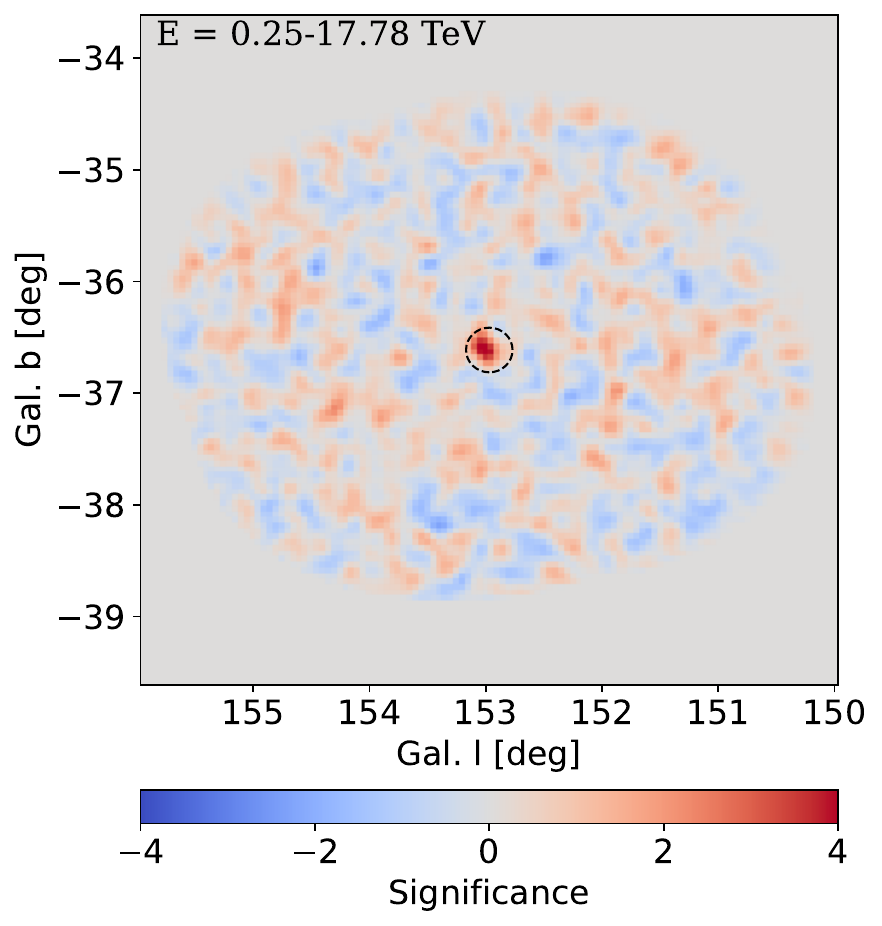}
\includegraphics[width=0.50\linewidth]{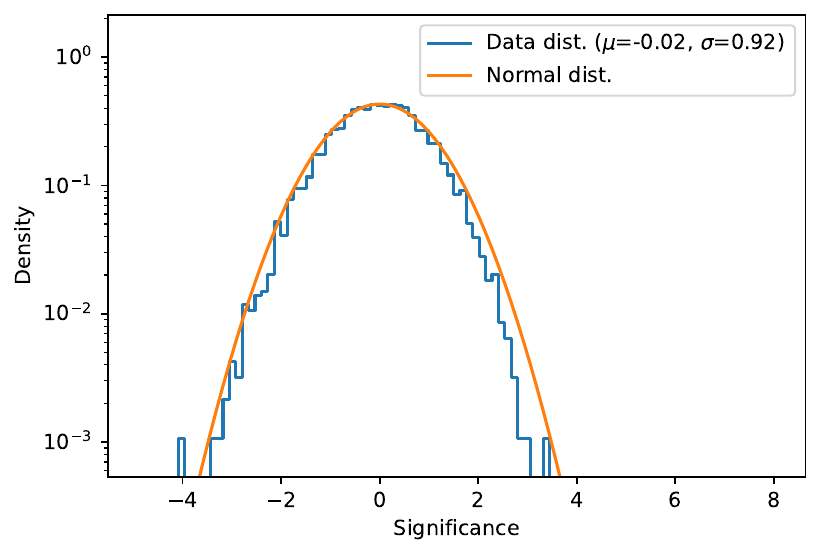} \\
\includegraphics[width=0.32\linewidth]{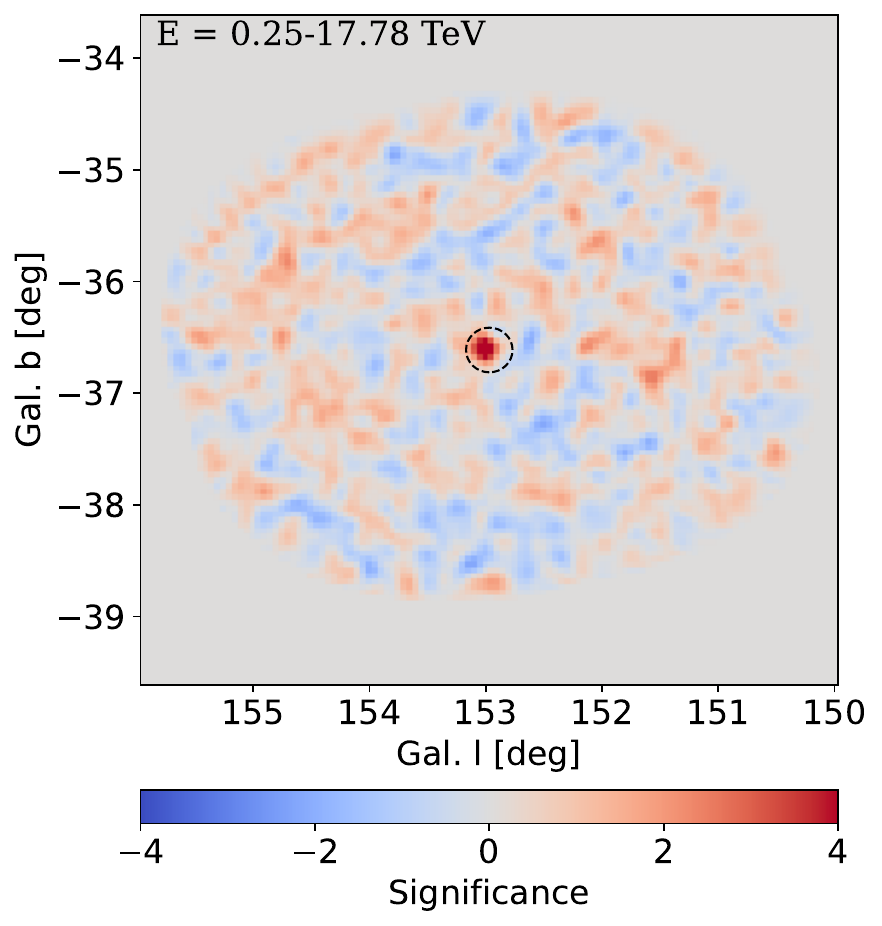}
\includegraphics[width=0.50\linewidth]{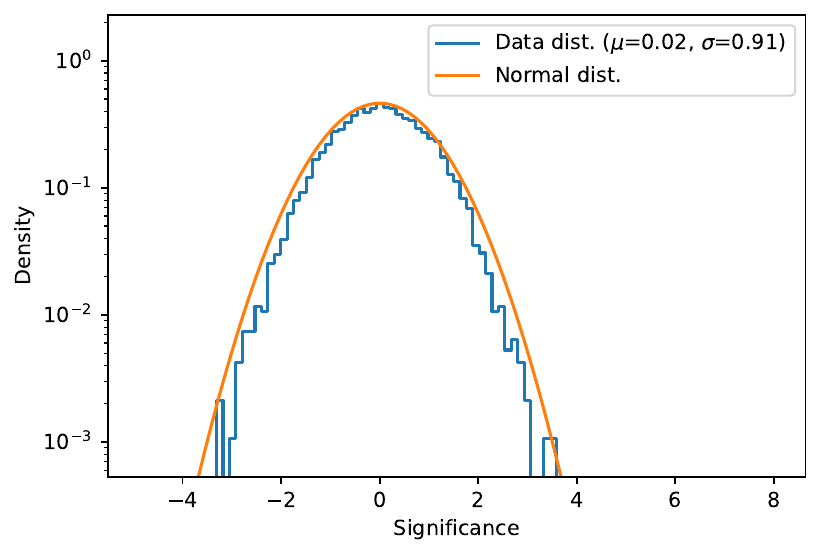}
\caption{
The significance maps and the significance distributions of $\gamma$-ray emissions of extragalactic observations of 1ES 0229+200 (observed at a mean elevation of $72^{\circ}$ and NSB of 6.0 D.C.) for a short exposure of 50 hours (top) and for a deep exposure of 220 hours (bottom).
The data significance distribution (blue curve) is sampled from outside of the source regions (highlighted by the black circles).
}
\label{fig:example_extragalactic_1ES0229_point_like}
\end{figure*}

\begin{figure*}
\centering
\includegraphics[width=0.32\linewidth]{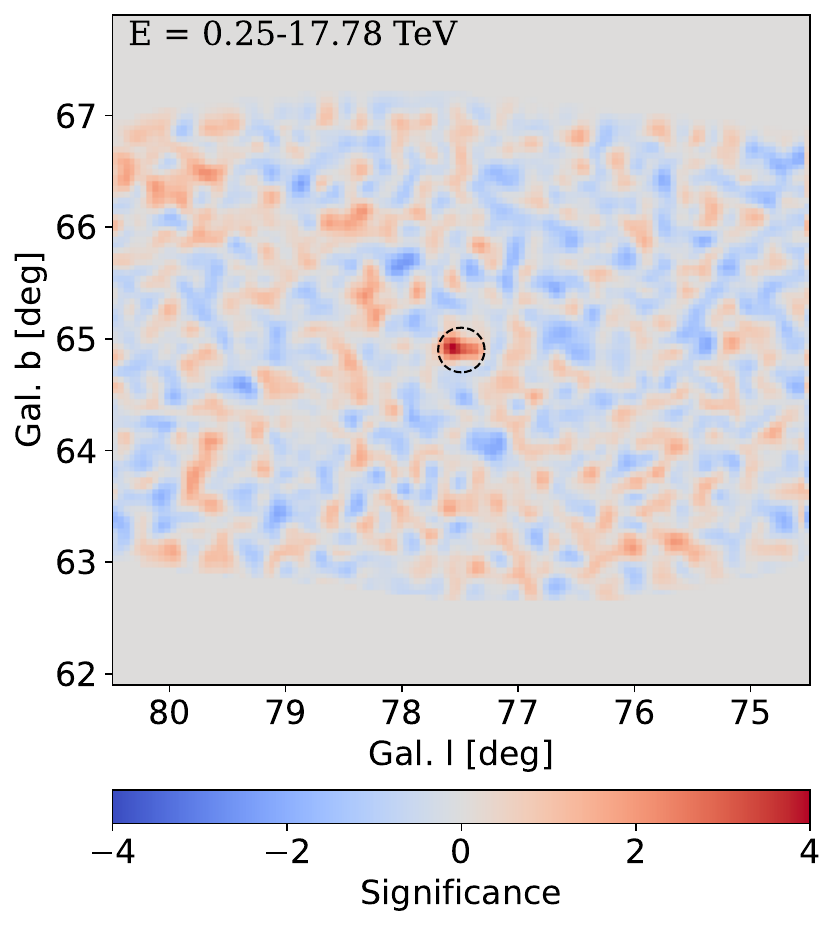}
\includegraphics[width=0.50\linewidth]{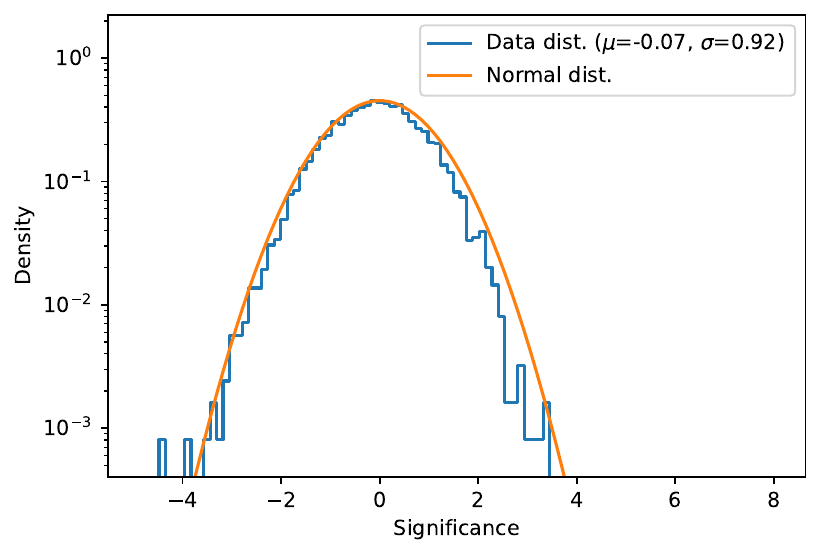} \\
\includegraphics[width=0.32\linewidth]{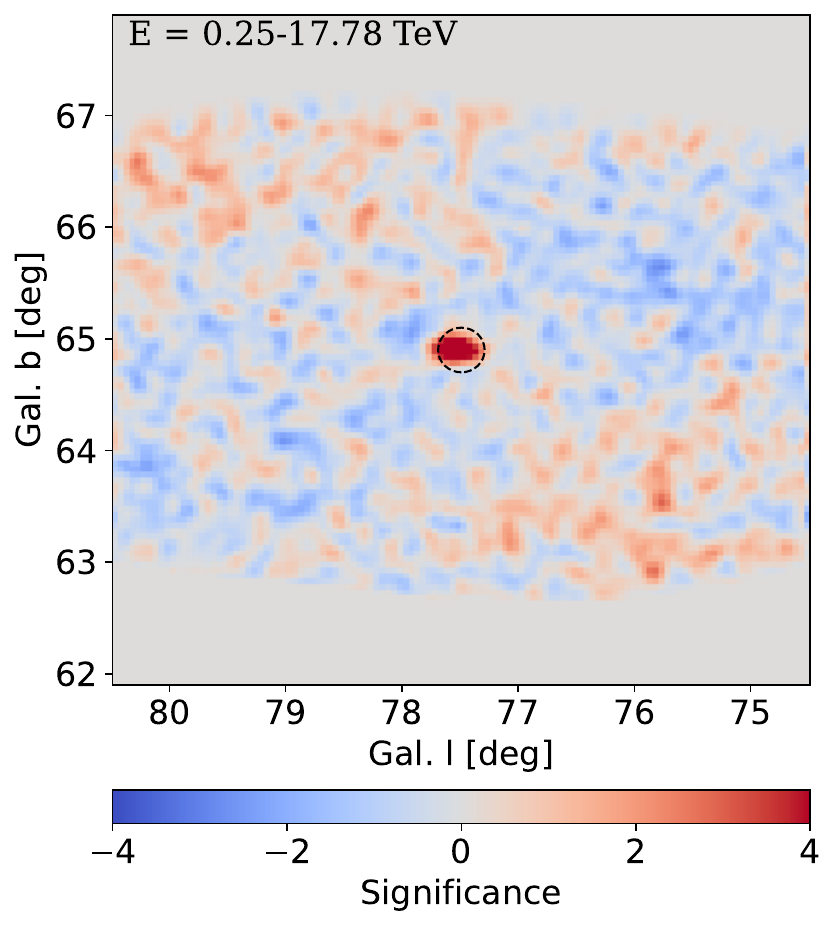}
\includegraphics[width=0.50\linewidth]{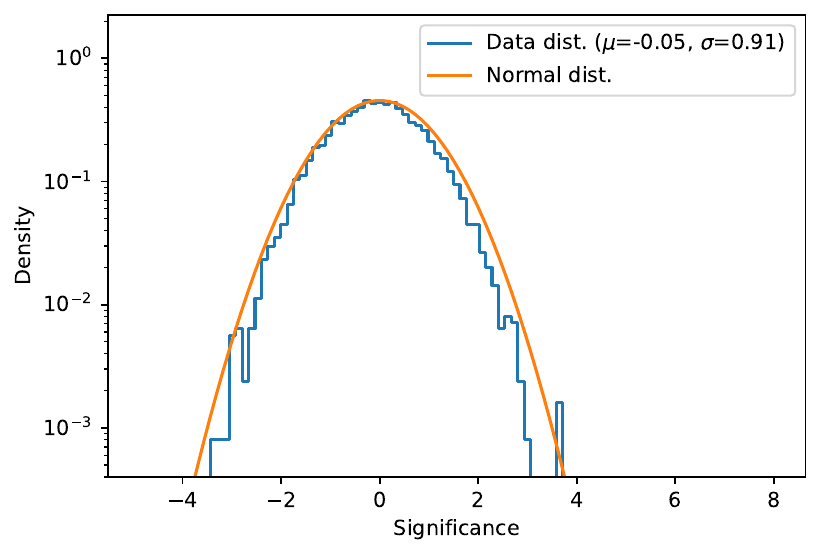}
\caption{
The significance maps and the significance distributions of $\gamma$-ray emissions of extragalactic observations of H 1426+428 (observed at a mean elevation of $73^{\circ}$ and NSB of 5.5 D.C.) for a short exposure of 50 hours (top) and for a deep exposure of 178 hours (bottom).
The data significance distribution (blue curve) is sampled from outside of the source regions (highlighted by the black circles).
}
\label{fig:example_extragalactic_H1426_point_like}
\end{figure*}

We start with two known point-like extragalactic sources in Figure \ref{fig:example_extragalactic_1ES0229_point_like} and \ref{fig:example_extragalactic_H1426_point_like}, which show the significance maps for the observations of 1ES 0229+200 (220-hour exposure) and H 1426+428 (178-hour exposure).
The singular templates are constructed from observations of multiple other point-like extragalactic sources similar to these two sources.

In the top panels of Figure \ref{fig:example_extragalactic_1ES0229_point_like} and \ref{fig:example_extragalactic_H1426_point_like}, we show the significance sky maps of the two sources (left panels) with short exposures of 50 hours, and we see that the sky maps show uniform random fluctuations outside of the source regions (black circles), and the corresponding significance distributions (sampled from outside of the source regions) also follow a normal distribution (right panels).

In the bottom panels of Figure \ref{fig:example_extragalactic_1ES0229_point_like} and \ref{fig:example_extragalactic_H1426_point_like}, we increased the exposures to 220 hours (1ES 0229+200) and 178 hours (H 1426+428), and we start to see non-uniform fluctuations outside of the source regions.
This is because the increase of exposures reduces the statistical fluctuations and reveals the systematic errors (which do not vary with the increase of exposures).
The estimated sizes of the systematic errors are included in the significance calculation, as described by Equation \ref{eq:updated_significance}, so the reported significance will not be inflated by systematic errors. 
Indeed, we see that the bulk significance distributions of the deep-exposure maps still agree with the normal distribution.

\begin{figure*}
\centering
\includegraphics[width=0.32\linewidth]{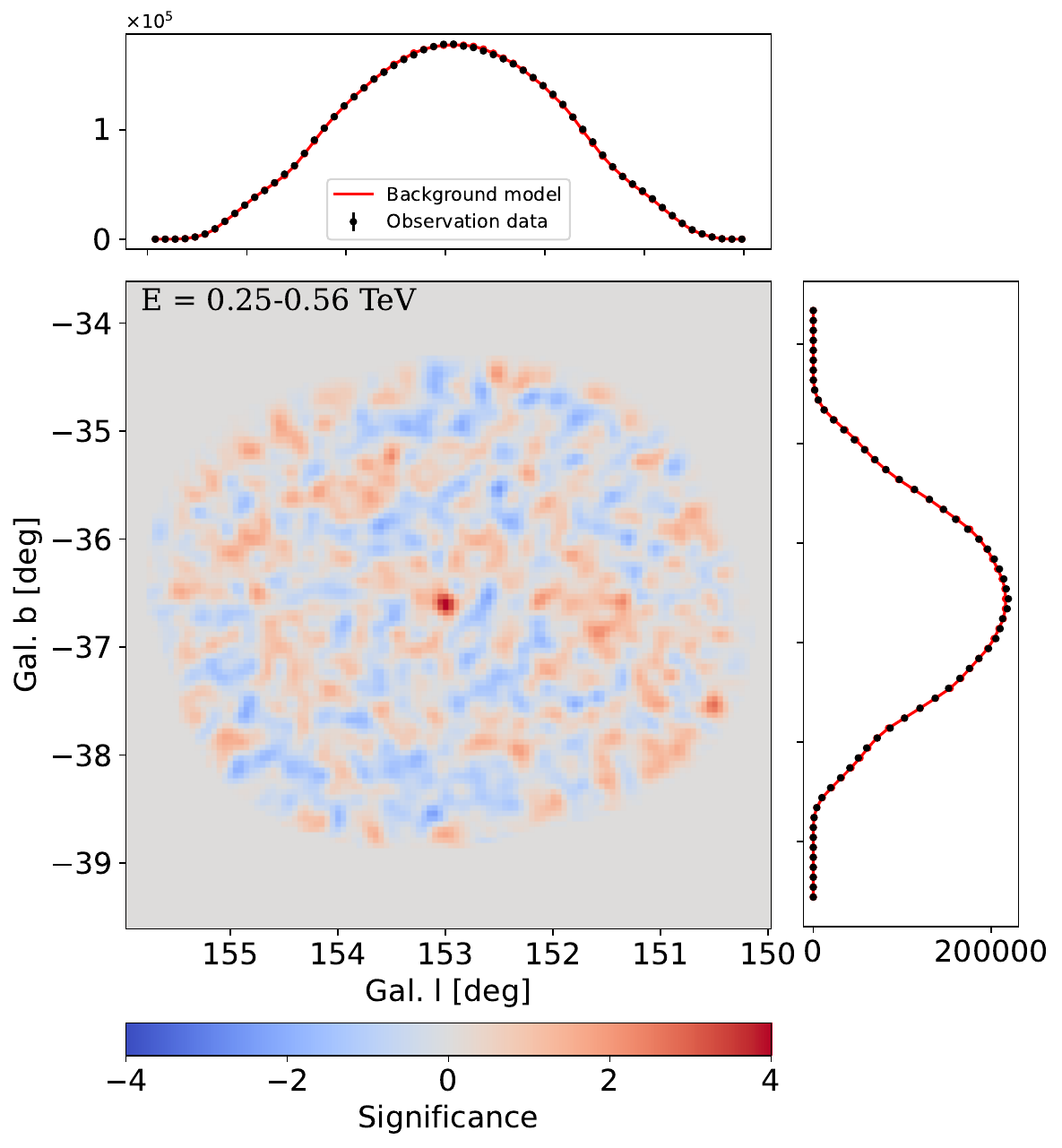}
\includegraphics[width=0.32\linewidth]{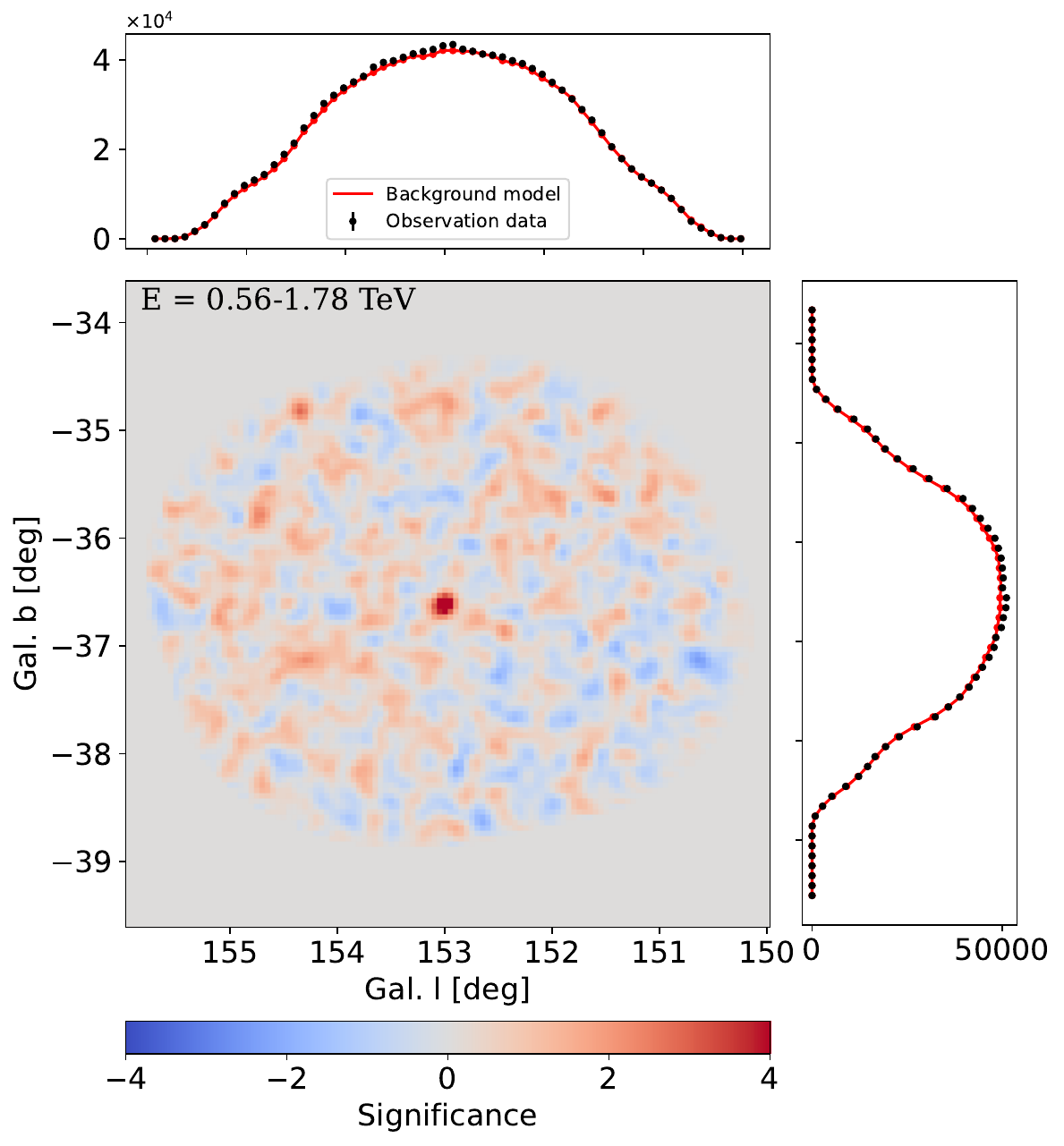}
\includegraphics[width=0.32\linewidth]{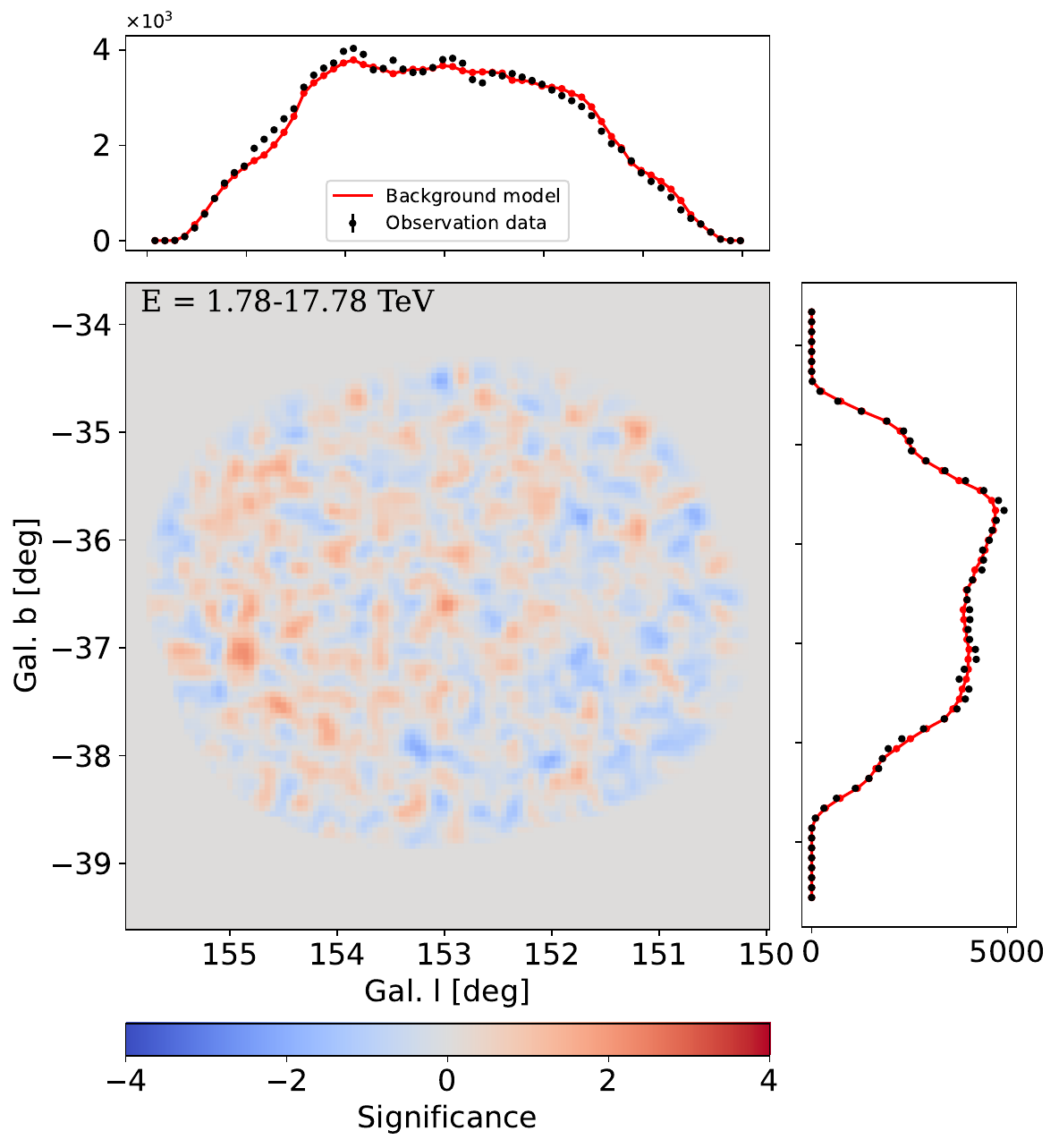}
\caption{
The energy-dependent significance maps and the projected 1-D distributions of the background model (red curves) and data events (black points) of $\gamma$-ray emissions of extragalactic observations of 1ES 0229+200 (a deep exposure of 220 hours observed at mean elevation of $72^{\circ}$ and NSB of 6.0 D.C.) in the $\left[0.25,0.56\right]$ (left), $\left[0.56,1.78\right]$ (middle), and $\left[1.78,17.78\right]$ (right) GeV ranges.
}
\label{fig:example_extragalactic_1ES0229_energy_dep}
\end{figure*}

We demonstrate the method's ability in providing energy-dependent background model in Figure \ref{fig:example_extragalactic_1ES0229_energy_dep}, which shows the energy-dependent significance sky maps and the projected 1-D event distributions of 1ES 0229+200 in the $\left[0.25,0.56\right]$, $\left[0.56,1.78\right]$ and $\left[1.78,17.78\right]$ GeV ranges.
From the 1D distributions of the data events, we can see that the background distribution changes dramatically from the low-energy range to the high-energy range, and the energy-dependent background shape is well-modeled by the STOICS method.

\subsection{Galactic point-like sources}

\begin{figure*}
\centering
\includegraphics[width=0.32\linewidth]{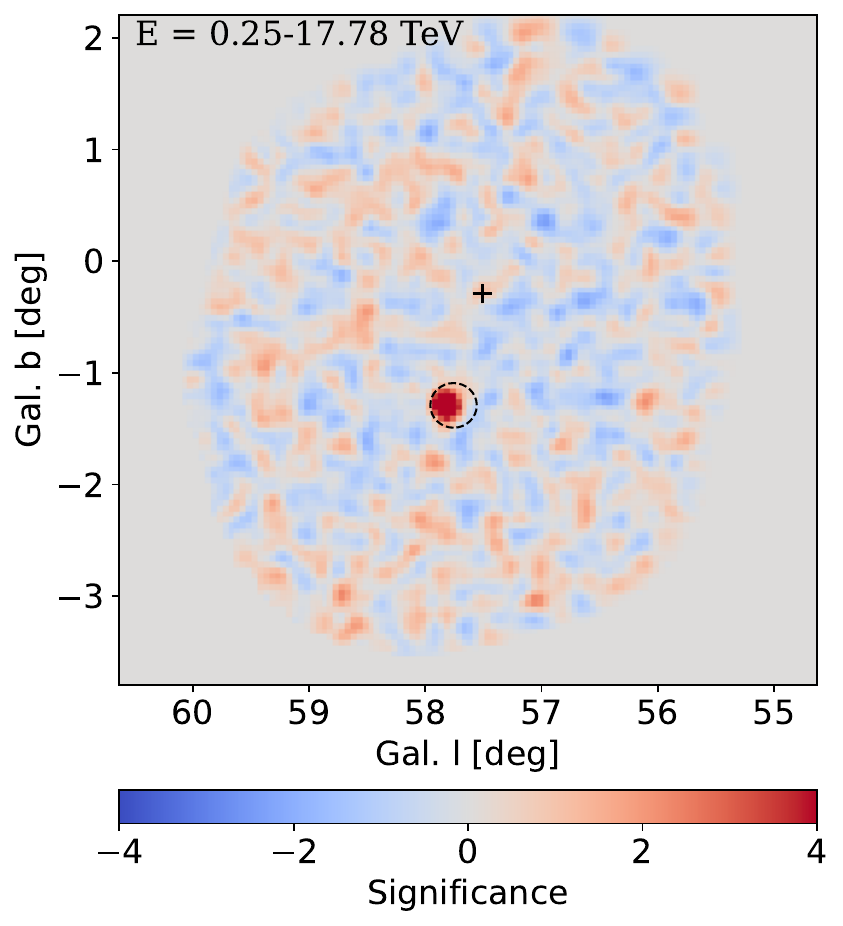}
\includegraphics[width=0.50\linewidth]{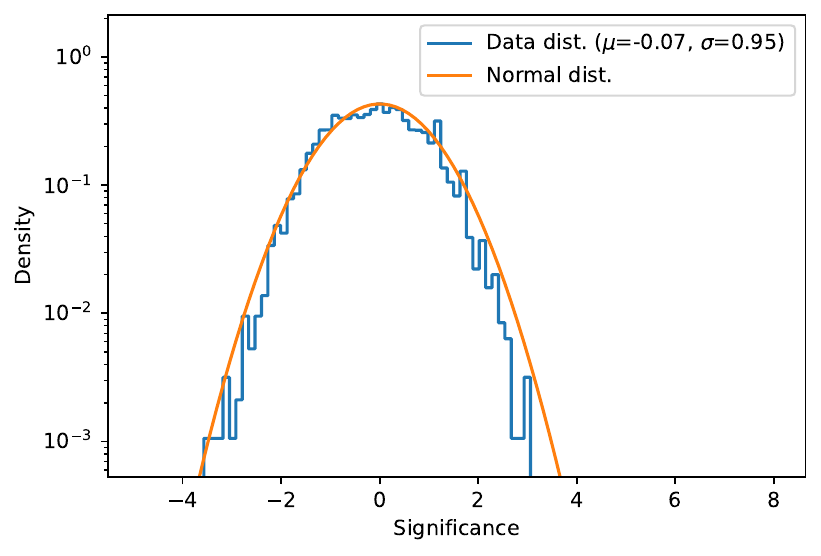}
\caption{
The energy-inclusive significance map (left) and the significance distribution (right) of $\gamma$-ray emissions of Galactic observations of HESS J1943+213 (48-hours exposure observed at a mean elevation of $72^{\circ}$ and NSB of 6.3 D.C.).
The significance distribution is sampled from outside of the source regions (highlighted by the black circles).
}
\label{fig:example_galactic_HESSJ1943_point_like}
\end{figure*}

\begin{figure*}
\centering
\includegraphics[width=0.32\linewidth]{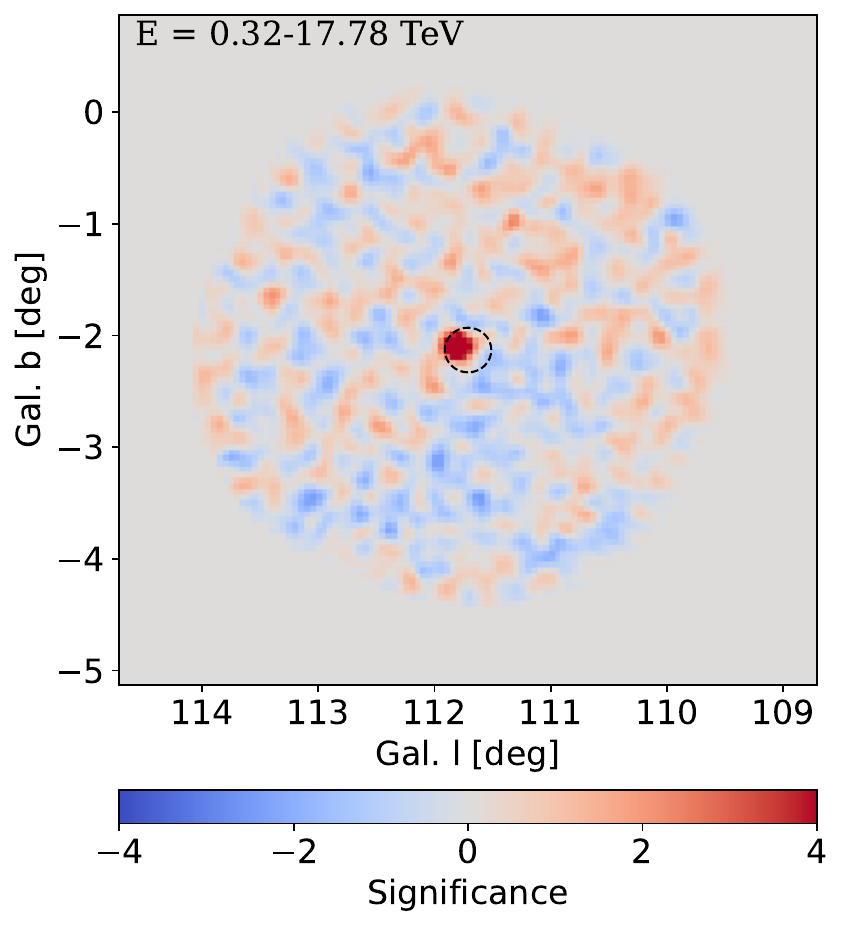}
\includegraphics[width=0.50\linewidth]{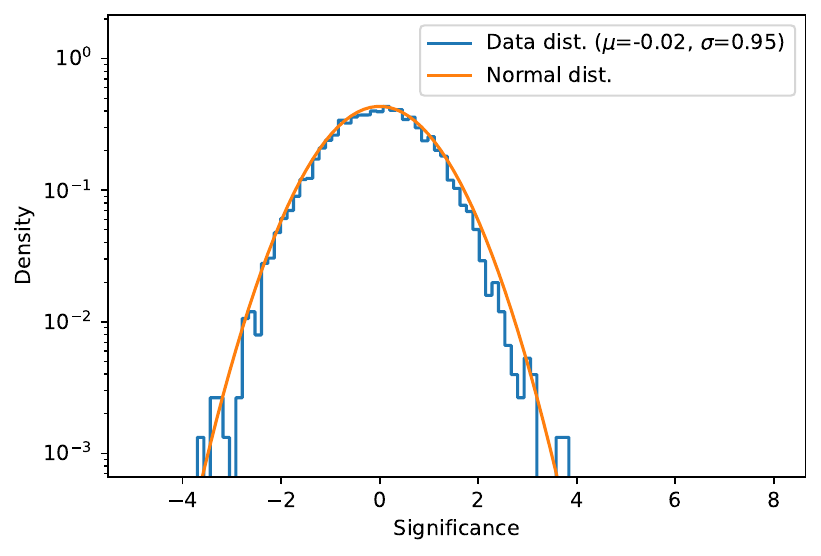}
\caption{
The energy-inclusive significance map (top-left) and the significance distribution (top-right) of $\gamma$-ray emissions of Galactic observations of Cas A (70-hours exposure observed at a mean elevation of $57^{\circ}$ and NSB of 6.5 D.C.).
The data significance distribution (blue curve) is sampled from outside of the source regions (highlighted by the black circles).
}
\label{fig:example_galactic_CasA_point_like}
\end{figure*}

We then show the significance maps for the observations of two point-like sources in the Galactic plane, HESS J1943+213 (48-hour exposure) and Cas A (70-hour exposure), in Figure \ref{fig:example_galactic_HESSJ1943_point_like} and \ref{fig:example_galactic_CasA_point_like}.

These Galactic-plane observations generally yield higher night-sky background than the extragalactic observations, and
these examples demonstrate that the STOICS method, with the singular templates constructed from extragalactic observations, is capable of modeling the backgrounds for these Galactic-plane observations without showing false-positive beyond the estimated systematic uncertainty.

\subsection{Galactic extended sources}

\begin{figure*}
\centering
\includegraphics[width=0.32\linewidth]{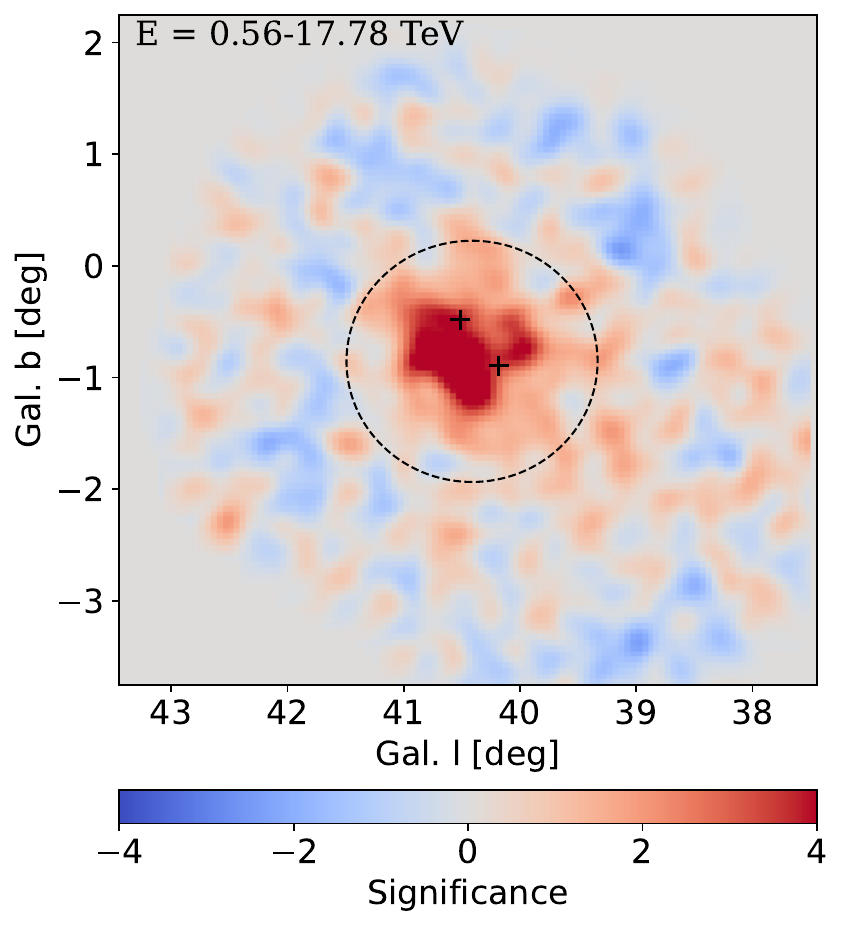}
\includegraphics[width=0.50\linewidth]{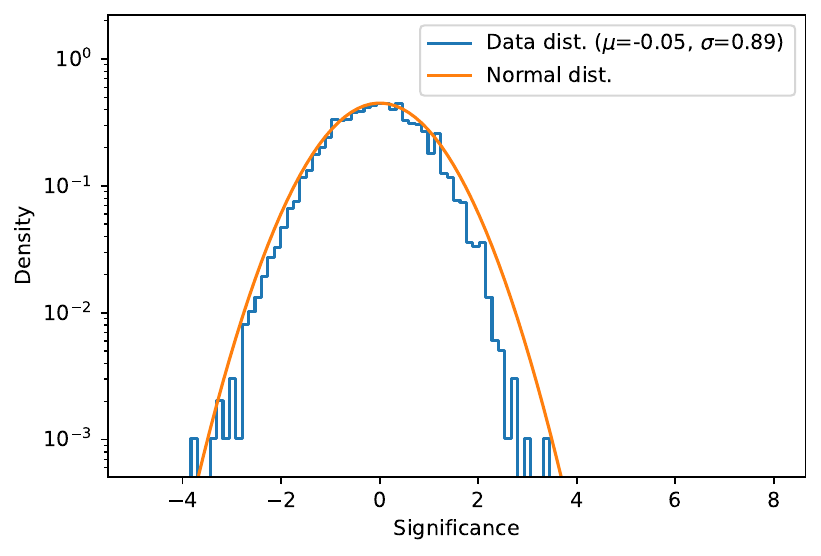}\\
\includegraphics[width=0.32\linewidth]{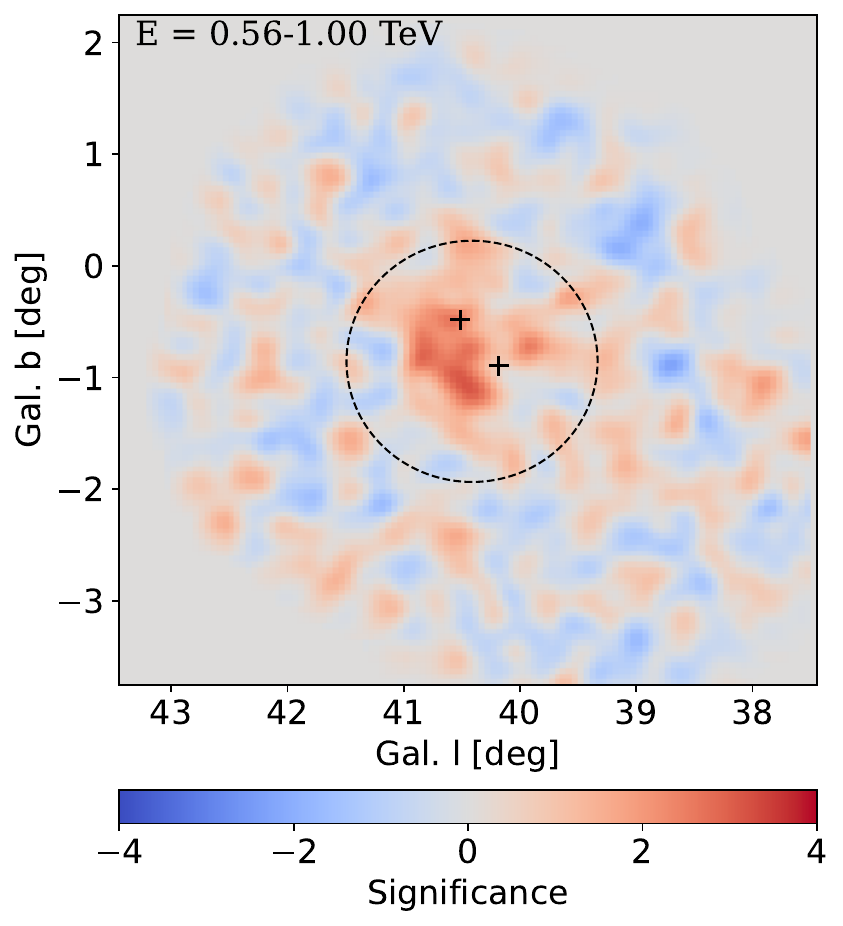}
\includegraphics[width=0.32\linewidth]{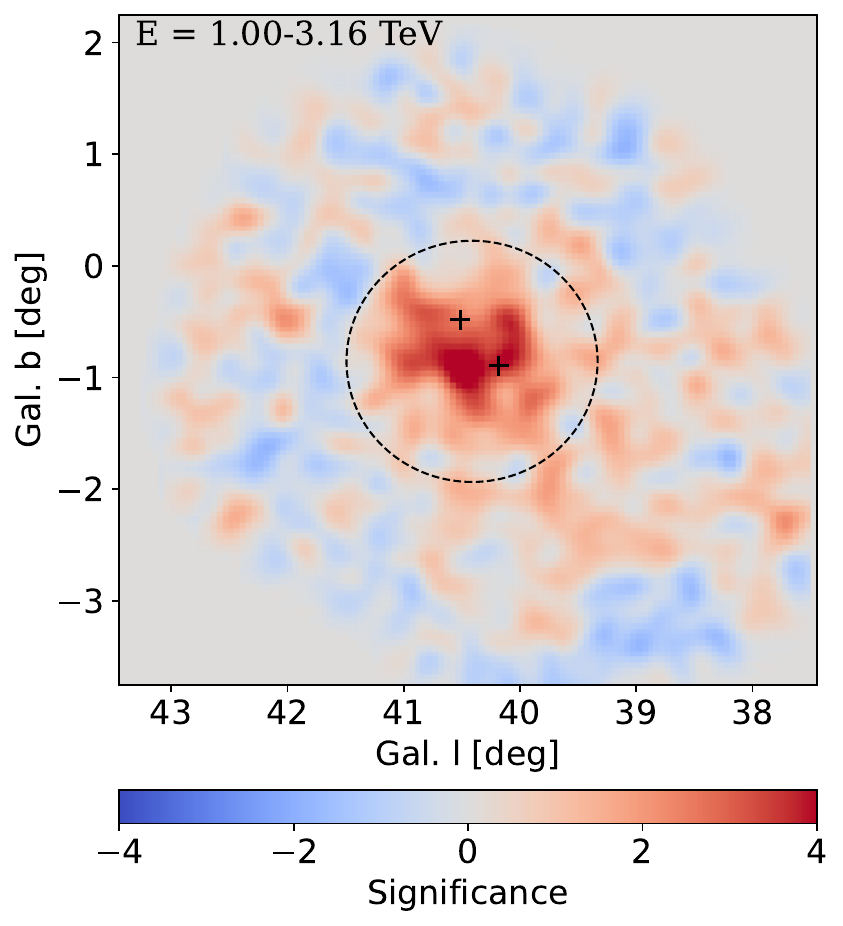}
\includegraphics[width=0.32\linewidth]{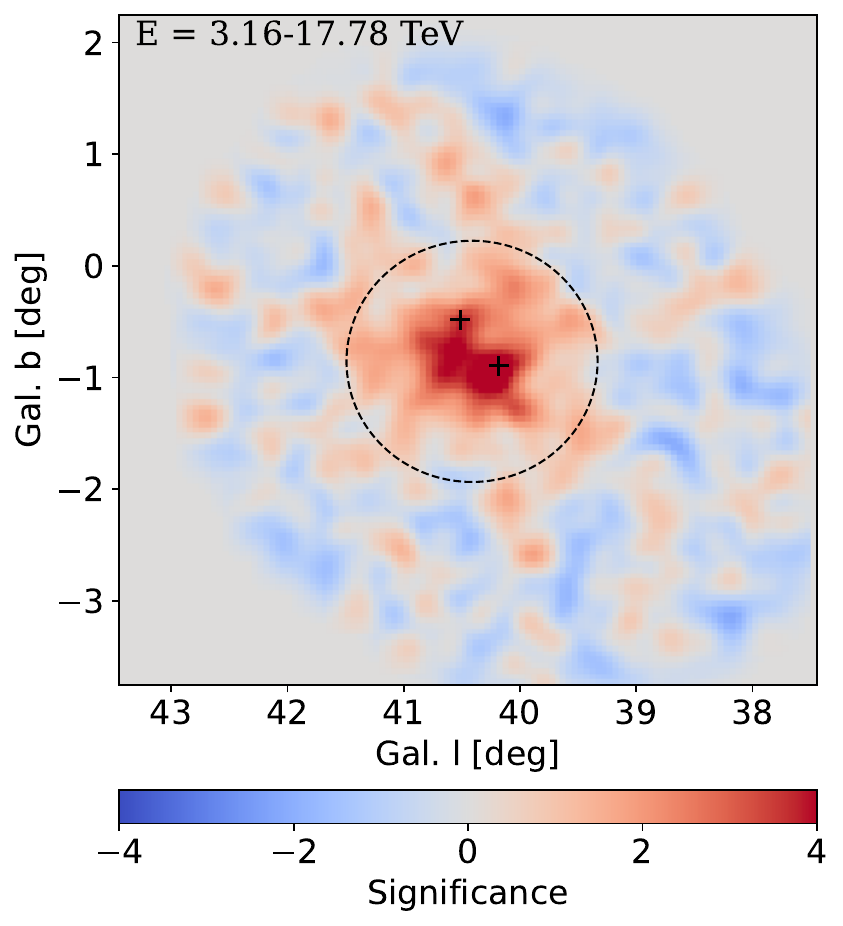}
\caption{
The significance maps of Galactic observations of MGRO J1908+06 (top-left, 159-hours exposure observed at a mean elevation of $58^{\circ}$ and NSB of 6.0 D.C.) and the data significance distribution (top-right, blue curve) sampled from outside of the source regions (highlighted by the black dashed circles).
The bottom panels show the energy-dependent significance maps in the $\left[0.56,1.00\right]$, $\left[1.00,3.16\right]$, and $\left[3.16,17.78\right]$ GeV ranges.
The black crosses indicate the locations of SNR G40.5-0.5 (upper-left) and PSR J1907+0602 (lower-right).
The black dashed circles indicate the source extensions ($3\times r_{39}$ of the 39\% containment radius of the 2D-Gaussian model) reported in \cite{cao2024first}.
}
\label{fig:example_galactic_extended_mgroj1908}
\end{figure*}

\begin{figure*}
\centering
\includegraphics[width=0.32\linewidth]{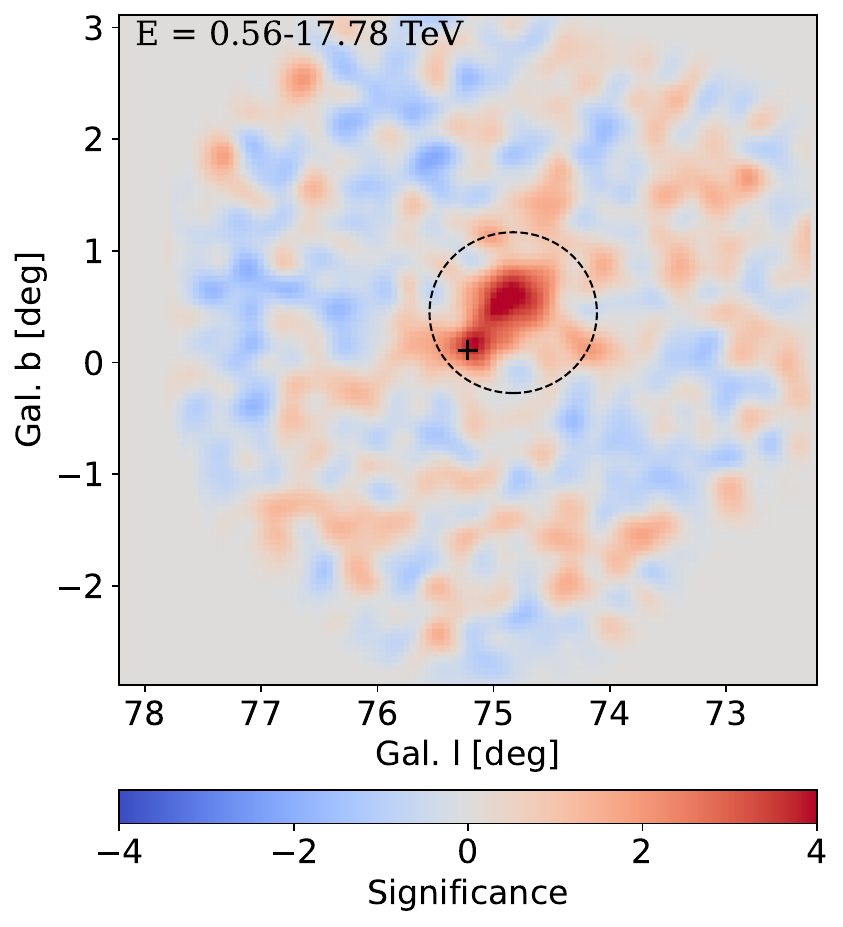}
\includegraphics[width=0.50\linewidth]{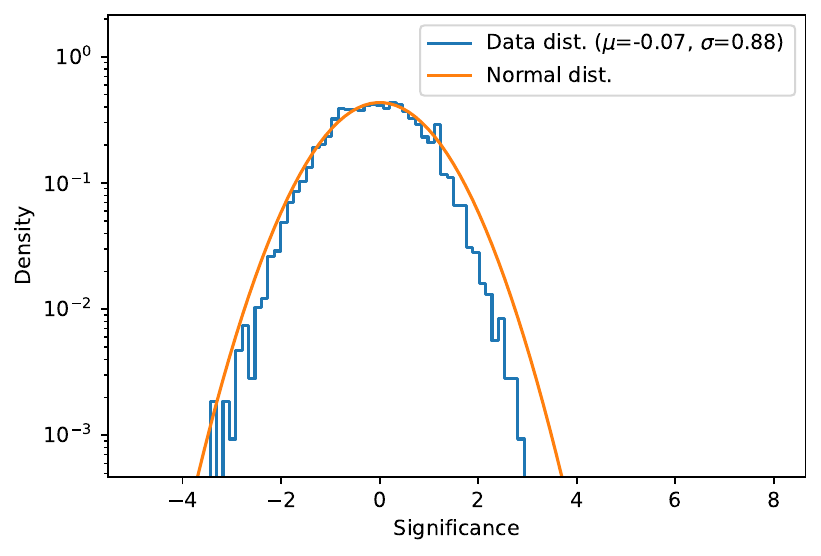}\\
\includegraphics[width=0.32\linewidth]{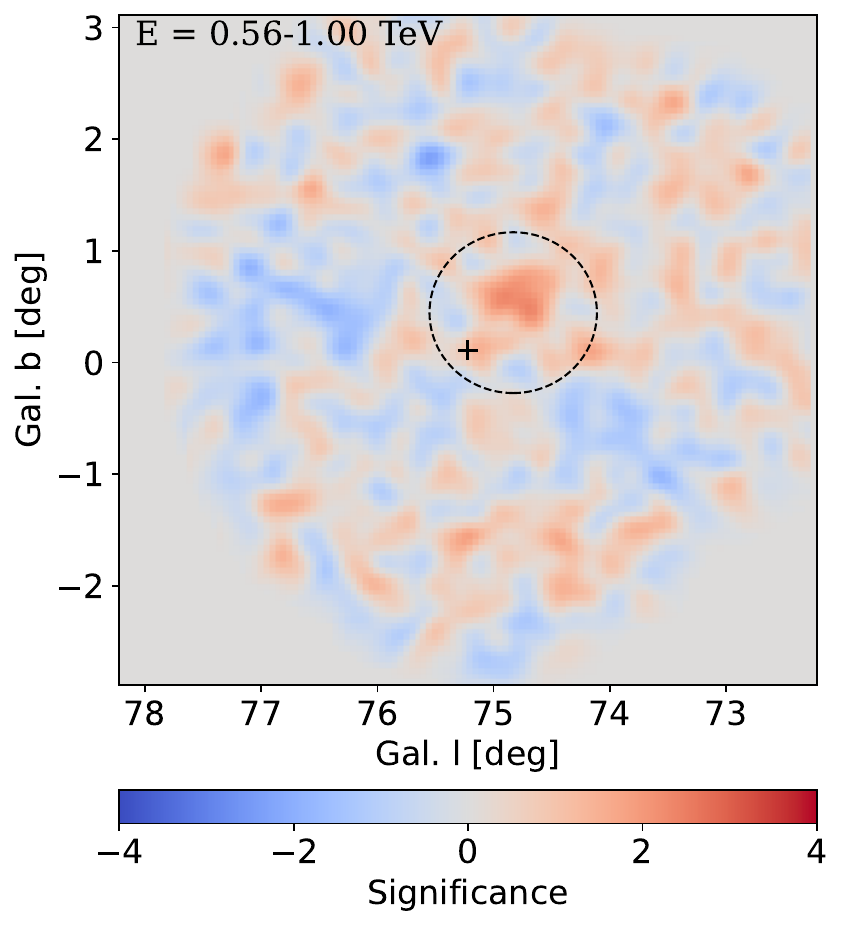}
\includegraphics[width=0.32\linewidth]{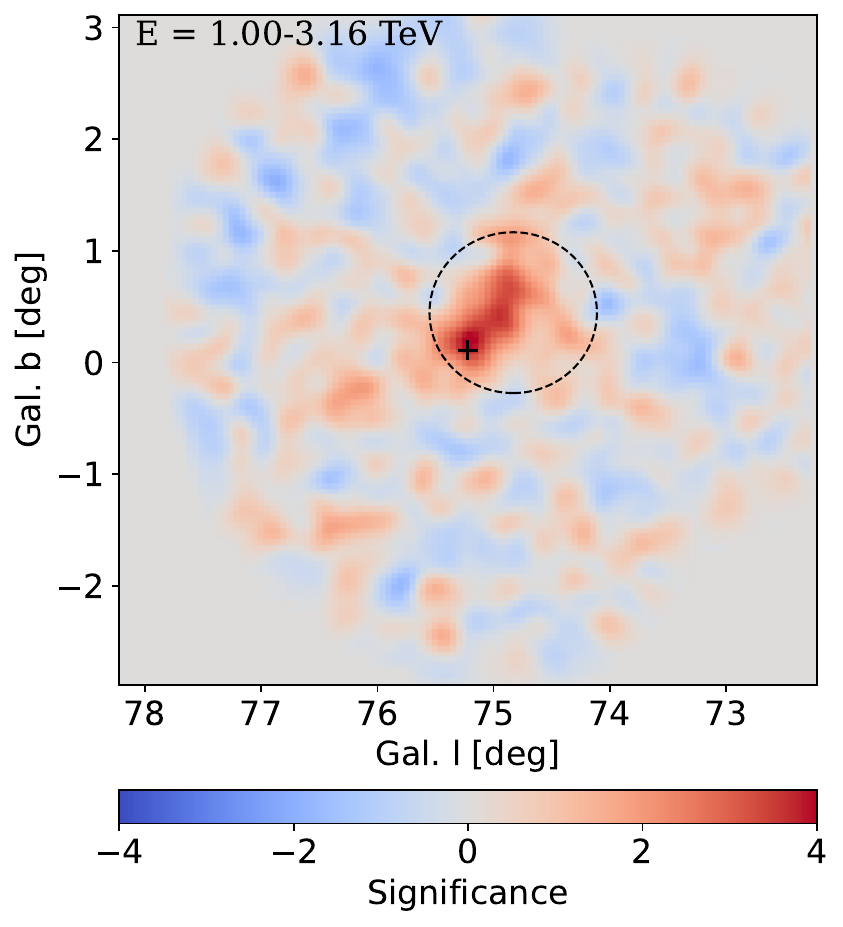}
\includegraphics[width=0.32\linewidth]{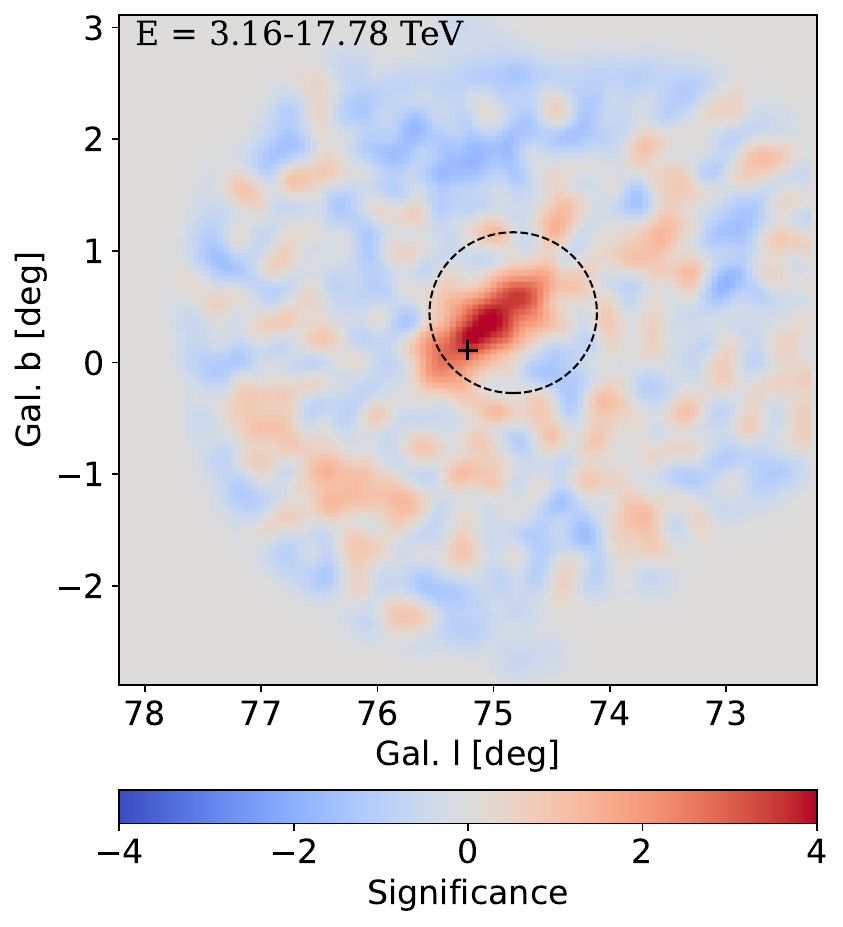}
\caption{
The significance maps of Galactic observations of MGRO J2019+37 (top-lft, 130-hours exposure observed at a mean elevation of $72^{\circ}$ and NSB of 6.6 D.C.) and the data significance distribution (top-right, blue curve) sampled from outside of the source regions (highlighted by the black dashed circles).
The bottom panels show the energy-dependent significance maps in the $\left[0.56,1.00\right]$, $\left[1.00,3.16\right]$, and $\left[3.16,17.78\right]$ GeV ranges.
The black cross indicates the locations of PSR J2021+3651.
The black dashed circles indicate the source extensions ($2\times r_{39}$ of the 39\% containment radius of the 2D-Gaussian model) reported in \cite{cao2024first}.
}
\label{fig:example_galactic_extended_dragonfly}
\end{figure*}

Finally, Figure \ref{fig:example_galactic_extended_mgroj1908} and \ref{fig:example_galactic_extended_dragonfly} demonstrate the ability of the STOICS method to detect extended $\gamma$-ray sources.
Analyses on these extended sources were reported previously by VERITAS and have angular extensions greater than $1^{\circ}$.

Figure \ref{fig:example_galactic_extended_mgroj1908} shows the significance map of MGRO J1908+06 (159-hour exposure) observed by VERITAS and analyzed with the STOICS method. 
MGRO J1908+06 is a bright extended gamma-ray source detected in the TeV energy range, first identified by the Milagro air shower array. Located in the Galactic plane, it is one of the most luminous sources observed by ground-based gamma-ray telescopes, including H.E.S.S., VERITAS, and HAWC. 
Figure \ref{fig:example_galactic_extended_dragonfly} shows the significance map of MGRO J2019+37 (130-hour exposure).
The source is located in the Galactic plane, near the Cygnus region, one of the most active star-forming regions in the Milky Way, and is regarded a key target for studying particle acceleration in astrophysical environments.

In both sky maps, we overlay the source extensions ($3\times r_{39}$) reported by \cite{cao2024first} indicated by the dashed circles. 
These results are consistent with our previous findings \citep{acharyya2024multiwavelength, aliu2014spatially}.

\section{Future application}

Cherenkov Telescope Array Observatory (CTAO) \citep{cta2018science} is a next-generation IACT that will operate as an open \emph{observatory}, where
science users can propose observations and have access to data. Given the
complexity of the instrument and IACT data analysis in general, CTAO will
provide user data products that have been pre-processed to a ``science-ready''
level that include fully calibrated \emph{event lists}, containing the reconstructed
energy, sky direction, arrival time, and other parameters of each $\gamma$-like
event, along with associated \emph{instrument response functions} (IRFs) needed for
scientific analysis. The IRFs allow for the transformation from physical fluxes
in true spectral and spatial coordinates to detected $\gamma$-ray counts in
reconstructed spatial and spectral coordinates, and contain a component that
models the spatial and spectral form of the residual background.

The instrumental response varies with pre-processing software configuration,
position in the field-of-view, $\gamma$-like energy range, site and sub-array
choice, pointing direction, atmosphere density, atmosphere quality, night-sky
optical background light levels, and instrumental degradation. While some
components of the IRFs can be computed from Monte-Carlo simulations of
$\gamma$-rays traversing the atmosphere and instrument, the residual background
component that models instrumental background after hadron rejection is one of
the most difficult to compute a priori due to uncertainties in the knowledge of
hadron physics \citep{ohishi2021effect,leitgeb2025probing}. 
In currently operating IACTs, the background models are computed
from existing source-free observations and are provided only as un-normalized spatial-spectral models. These models must therefore be normalized by the user
analyzing the data using control regions in the field of view, which requires
prior knowledge of all $\gamma$-ray sources in the field of view, or by matching
observations taken at different times, which can lead to high systematic errors.  
Furthermore, to reduce the amount of source-free data needed to build such a model, simplifying assumptions like radial symmetry in the field-of-view are sometimes made, which can lead to high systematic errors in the case of fields where that assumption fails, such as for observations at large zenith angles (validated using observations taken at $>40^{\circ}$ elevation angles in this paper), in regions with non-uniform night-sky background light, or taken with asymmetric sub-arrays.

The technique described in this paper could potentially be useful for an observatory like CTAO for three reasons: 1) providing per-observation pre-normalized background models to users without prior knowledge of the science case, 2) providing a background model with two spatial dimensions that can easily account for field-of-view complexities and 3) allowing for the study of emission that is similar to or larger than the field-of-view of the instrument, which itself varies considerably with energy and sub-array choice. While the first two make science analysis easier for users and lower systematics, the latter is critical for certain CTAO
science cases such as indirect dark matter detection \cite{acharyya2024indirect, ahnen2018indirect, abdallah2020search}, large-scale
galactic outflows \cite{ackermann2014spectrum}, galactic diffuse emission \cite{ackermann2012fermi}, and
TeV halos \cite{di2019detection,aharonian2023detection,abeysekara2017extended}. 
These specific cases require
understanding the residual background across the full field-of-view and energy range of the instrument at all observation times, with few or no off-source
regions possible for normalizing a model to observed data.

\section{Conclusion}

This paper describes the STOICS method, a novel analysis technique for IACT observations in the case when the source angular extension exceeds or occupies a significant part of the field-of-view.
We showed that using the CR-like events (events that failed the $\gamma$/hadron separation cut) and the SVD analysis of the air-shower event distributions in the $\gamma$-ray camera frame, the method is capable of providing reliable background modeling for observations over a wide range of observing conditions given a loose match criterion for OFF runs (defined in Section \ref{sec:eigen_method}).

We should note a limitation of STOICS method is that the method is prone to overestimating the background in the presence of a strong $\gamma$-ray source because the signal events can spill into the CR-like region.
The method requires the CR-like region to be $\gamma$-ray-free in order to make a background prediction in the $\gamma$-like region, and this means that the cuts on $\gamma$/hadron separation variables need to be relaxed in the presence of a strong source.

This novel background estimation technique is designed for IACT analyses dealing with extended sources or sources with poor knowledge of localization.
This method allows reliable energy-dependent morphology measurements for sources with large angular extents, which will be helpful in the era of CTAO in reaching its full potential capacity of wide field-of-view observations.

\section*{Acknowledgment}
RS thanks the VERITAS Collaboration for providing their internal analysis software package and their archival data. 
RS gratefully acknowledges Vladimir Vassiliev for his guidance. 
RS thanks Reshmi Mukherjee for her advice and overall support. 
RS was supported by NSF grants PHY-2110497 and PHY-2411023 at Barnard College.


\bibliographystyle{cas-model2-names}

\bibliography{cas-refs}



\end{document}